\documentclass[journal]{IEEEtran}
\usepackage{amsmath,amsfonts,amssymb}
\usepackage{array}
\usepackage{textcomp}
\usepackage{stfloats}
\usepackage{url}
\usepackage{verbatim}
\usepackage{graphicx}
\usepackage[caption=false,font=footnotesize]{subfig}
\usepackage{tabularx}
\usepackage{gensymb}  
\usepackage{enumitem}
\usepackage{booktabs}
\usepackage{algorithm}
\usepackage{algpseudocode}  
\usepackage{xfrac}
\usepackage[colorlinks=true, linkcolor=blue, citecolor=blue, urlcolor=blue]{hyperref}
\usepackage{cite}
\usepackage{xcolor}
\usepackage{soul}

\begin{document}
\bstctlcite{IEEEexample:BSTcontrol}

\title{Model-Based Beam-Steered Optical Wireless Positioning with Single-LED Single-Photodiode for 3D Localization}
\author{Kevin Acuna-Condori, Bastien Béchadergue, Hongyu Guan, and Luc Chassagne
\thanks{Manuscript received MM DD, 2025.
This work was supported in part by the French National Research Agency (ANR) through the Project SAFELiFi under Grant ANR-21-CE25-0001-01, and in part by the European Union's Smart Networks and Services Joint Undertaking (SNS JU) under grant agreement No. 101139292 (\textit{Corresponding author: Kevin Acuna-Condori.})
}
\thanks{
The authors are with the Université Paris-Saclay, UVSQ, LISV, 78140, Vélizy-Villacoublay, France (e-mail: kevin-jose.acuna-condori@uvsq.fr).
}
}
\markboth{IEEE Transactions on Communications,~Vol.~XX, No.~X, 2025}%
{Acuna-Condori \MakeLowercase{\textit{et al.}}: Model-Based Beam-Steered OWP with Single-LED Single-PD for 3D Localization}
\maketitle

\begin{abstract}
State-of-the-art optical wireless positioning (OWP) commonly reaches centimeter-level accuracy by depending on dense multi-light-emitting diodes (LED) infrastructures, photodiode (PD) arrays, or image-sensor receivers, incurring hardware complexity and deployment cost. 
This paper introduces a single beam-steered LED, single-PD OWP architecture that achieves three-dimensional (3D) localization without cameras or PD arrays. The core idea is to steer the transmitter through $K$ known orientations and exploit the resulting received signal strength variations at the PD to first estimate the LED-to-PD direction, and then recover the distance from a single cooperative beam-aligned measurement. We derive a direction error bound (DEB) for the direction-finding stage and a position error bound for the complete 3D positioning system, and cast the steering-pattern design as a genetic algorithm that minimizes the DEB over a 3D testbed. We develop three direction estimators: a statistically efficient closed-form generalized least squares (GLS), a lightweight weighted least squares (WLS) approximation, and an iterative nonlinear least-squares (NLS) estimator. We prove that all three estimators are mathematically independent of the receiver orientation, enabling direction finding without receiver pose knowledge or control. Simulations over 1,792 testbed positions with 1,000 Monte Carlo trials demonstrate, for $K=5$ orientations, sub-degree mean direction-finding error (GLS: $0.54^\circ$, WLS: $0.61^\circ$, NLS: $0.52^\circ$) and centimeter-level average 3D positioning error (GLS: 2.00\,cm, WLS: 2.25\,cm, NLS: 1.90\,cm), with robustness to random receiver tilts (degradation below 3\%).
\end{abstract}


\section{Introduction}

\IEEEPARstart{G}{lobal} navigation satellite systems (GNSS) provide ubiquitous outdoor positioning but are unreliable indoors due to severe attenuation, blockage, and multipath. Dedicated indoor positioning systems are therefore required, and radio-frequency (RF) techniques leveraging existing communications infrastructure---such as Wi-Fi \cite{He2016}, Bluetooth Low Energy (BLE) \cite{Faragher2015}, ultra-wideband (UWB) \cite{Alarifi2016}, radio-frequency identification (RFID) \cite{Ni2004}, and emerging 5G positioning \cite{Italiano2025}---are widely studied and deployed \cite{Zafari2019, Italiano2025}. However, in cluttered interiors they are strongly affected by multipath and time-varying interference in shared spectrum, typically yielding meter-level accuracy unless dense infrastructure, calibration, or specialized hardware is used \cite{Lymberopoulos2015, Leitch2023}.

Optical wireless positioning (OWP) has emerged as a compelling complement that exploits the favorable propagation properties of light for indoor localization \cite{Wang2024b}. The spatial confinement of optical signals (no wall penetration), immunity to RF electromagnetic interference, high spatial reuse, and the deterministic nature of line-of-sight (LOS) channels enable centimeter-level accuracy with cost-effective light-emitting diodes (LED) \cite{Trevlakis2023}. Moreover, the same LED infrastructure can simultaneously provide optical wireless communication (OWC)---often termed light-fidelity (LiFi) \cite{Haas:16}---so that illumination, data, and positioning coexist in a unified platform. These attributes make OWP attractive for navigation and asset tracking in malls, airports, and museums where GNSS is unavailable, as well as in electromagnetic-sensitive environments such as hospitals and airplane cabins \cite{Wang2024b, Zhu2025}.

Within OWP, two estimation paradigms coexist. Data-driven methods---including fingerprinting with support vector machines, random forests, multilayer perceptrons, convolutional neural networks, recurrent models, and Transformer architectures \cite{Rekkas2023, Tran2022}---can be effective given abundant labeled data, but are often opaque, sensitive to domain shift (hardware changes, illumination variations, occlusion), and require periodic retraining, offering limited physical guarantees \cite{Glass2021}. By contrast, model-based estimators grounded in radiometry and geometry, e.g., by Lambertian propagation or field-of-view (FOV) constraints, provide interpretability, sample efficiency, principled uncertainty quantification, and support joint system co-design \cite{Deng2023}. This work adopts the model-based paradigm.

\begin{table*}[t]
\centering
\scriptsize
\setlength{\tabcolsep}{3pt} 
\renewcommand{\arraystretch}{1.15}
\caption{State-of-the-art single-LED OWP systems and reported performance.}
\label{tab:single_led_vlp}
\resizebox{\textwidth}{!}{
\begin{tabular}{|c|c|p{3.8cm}|c|c|p{3.8cm}|c|c|c|}
\hline
\textbf{Year} & \textbf{Work} & \textbf{Configuration} & \textbf{Rx Orientation} & \textbf{2D/3D} & \textbf{Method} & \textbf{Error (APE)} & \textbf{Room ($m^3$)} & \textbf{Notes} \\
\hline
2020 & \cite{Qin2020} & 1 LED / 4 PD (1 horizontal + 3 tilted) & Fixed & 3D & Model-based (Geometrical equations) & 2.52 cm & \(1 \times 1 \times 1.5\) & Sim. \\ \hline
2021 & \cite{Chen2021} & 1 LED / 3 tilted PD & Fixed & 2D & Long Short Term Memory- Fully Connected Network & 0.92 cm & \(1 \times 1.1 \times 1.75\) & Exp. \\ \hline
2021 & \cite{Qin2021} & 1 LED / 4 PD (1 horizontal + 3 tilted) & Fixed & 2D & Bayesian algorithm & 9.27 cm & \(3 \times 3 \times 3\) & Exp. \\ \hline
2022 & \cite{Liu2022} & 1 LED / 1 rotatable PD & Fixed & 2D & Extreme Learning Machine & 1.74 cm & \(4 \times 4 \times 3.1\) & Sim. \\ \hline
2023 & \cite{Zia-Ul-Mustafa2023} & 1 LED / 4 PD & Fixed & 2D & Least Square & 7 cm (CDF$_{90\%}$) & \(6 \times 6 \times 3\) & Exp. \\ \hline
2024 & \cite{Zhang2024} & 1 LED / 4 PD & Fixed & 2D & Extreme Learning Machine / WKNN & 2.93 cm & \(5 \times 5 \times 3\) & Sim. \\ \hline
2024 & \cite{Li2024} & 1 LED / 4 PD (1 horizontal + 3 tilted) & Fixed & 3D & Deep Residual Shrinkage method & 5.85 cm & \(3.6 \times 3.6 \times 3.0\) & Exp. \\ \hline
2024 & \cite{Wang2024} & 1 LED / 1 PD (tilted-rotatable PD) & Rotatable & 2D & Least Squares + Improved Grey Wolf Optimizer (LS-IGWO) & 1.65 cm & \(0.5 \times 0.5 \times 0.3\) & Exp. \\ \hline
2024 & \cite{Ma2024} & 1 LED / 4 PD + IMU (orientation) & Arbitrary & 3D & MultiLayer Perceptron & 1.77 cm & \(1.5 \times 1.5 \times 1.5\) & Exp. \\ \hline
2025 & \cite{Chassagne2025} & 1 reorientable LED / 1 PD & Arbitrary & 2D & Model-based (Power ratio) & 5.9 cm (CDF$_{90\%}$) & \(4 \times 4 \times 2.5\) & Sim. \\ \hline
2025 & \cite{Shi2025} & 1 LED / 1 rotatable PD & Fixed & 2D & Least Square & 4.96 cm & \(5 \times 5 \times 4\) & Exp. \\ \hline
2025 & \cite{ShiJin2025} & 1 LED / 16 IRS / 1 PD & Fixed & 2D & Levenberg--Marquardt algorithm & 5.6 cm (RMSE) & \(4 \times 5 \times 5\) & Sim. \\ \hline
2026 & Ours$^\dagger$ & 1 reorientable LED / 1 PD & \begin{tabular}[c]{@{}c@{}}Arbitrary \\ Controlled\end{tabular} & \begin{tabular}[c]{@{}c@{}}D-F \\ 3D\end{tabular} & Model-based (closed-form GLS) & \begin{tabular}[c]{@{}c@{}}$0.54^\circ$ \\ 2.00\,cm\end{tabular} & \(3 \times 3 \times 2\) & Sim. \\
\hline
\end{tabular}
}
\begin{flushleft}
\scriptsize \emph{Note:} ``Sim.'' denotes simulation results; ``Exp.'' denotes experimental results. $^\dagger$Two-stage system: direction finding (D-F) operates with any receiver orientation (arbitrary); 3D distance recovery requires a single cooperative PD reorientation toward the transmitter (controlled). See Section~\ref{subsec:nr_independence}.
\end{flushleft}
\end{table*}

The majority of OWP implementations follow a multi-LED paradigm, where several fixed LEDs with known positions act as anchors and the receiver resolves its position via triangulation or trilateration, analogous to GNSS \cite{Zhuang2018, Wang2024b}. When reusing visible-light luminaires---the standard approach in visible light positioning (VLP)---the anchor constellation is co-designed with illumination: placement, power, and beam patterns must satisfy lighting standards and uniformity, constraining anchor geometry and modulation \cite{photonics9100750}. Crucially, at least three to four LEDs must be concurrently visible to the receiver to resolve a three-dimensional (3D) position \cite{Plets2019, electronics8111311, ALMADANI2021126654, Zhang2020}, a condition that may not hold in environments with sparse lighting. Installing and maintaining a dense multi-LED network, with unique identifiers or modulation for each lamp, further increases deployment cost and complexity \cite{Cossu2022}. Consequently, reducing the infrastructure burden by localizing with a single light source is a highly desirable research goal \cite{Rekkas2023}.

Table~\ref{tab:single_led_vlp} summarizes representative single-LED OWP systems. The reported solutions adopt a variety of receiver architectures to compensate for the limited information provided by a single source at a fixed orientation: multiple tilted photodiodes (PD) that create directional diversity \cite{Qin2020, Qin2021, Zia-Ul-Mustafa2023, Zhang2024, Li2024}, a single mechanically rotatable PD \cite{Liu2022, Wang2024, Shi2025}, or aiding sensors such as inertial measurement units (IMU) \cite{Ma2024} and intelligent reflecting surfaces \cite{ShiJin2025}. Several systems are restricted to two-dimensional (2D) positioning \cite{Chen2021, Qin2021, Liu2022, Zia-Ul-Mustafa2023, Chassagne2025, Zhang2024, Wang2024, Shi2025, ShiJin2025}, whereas fewer achieve 3D localization \cite{Qin2020, Li2024, Ma2024}. A common thread emerges: with a single LED at a fixed orientation, one received signal strength (RSS) measurement is insufficient to resolve a 3D position; existing single-LED systems therefore compensate by placing hardware or computational complexity at the receiver side---through PD arrays, continuous mechanical rotation, or sensor fusion---limiting the simplicity and scalability of the user equipment.

Recent advances in OWC suggest a fundamentally different strategy: shifting the spatial diversity requirement from the receiver to the transmitter infrastructure through beam steering. In particular, microelectromechanical system (MEMS) micromirror gimbals redirect the LED beam mechanically, thereby preserving the native Lambertian emission pattern while achieving kHz-rate reorientation \cite{Liu:25}. Moreover, OWC roadmaps increasingly regard dynamic beam steering as a core enabler for high-capacity, mobile, multiuser links and blockage mitigation \cite{mi13070990, Zeng2023, Jin2023}. A recent comparative study has further shown that beam-steered transmitters substantially outperform fixed-antenna configurations in terms of indoor communication coverage \cite{AlSatai2026}, making it natural to exploit the same transmitter-side capability for positioning. In this work, we employ a near-infrared (NIR) LED transmitter, as opposed to conventional VLP which must respect visible-light illumination standards. Since NIR is imperceptible to humans, the positioning link is decoupled from room lighting constraints, allowing unrestricted steering and reorientation of the source \cite{Zhu2025}.

Building on these observations, we study a novel model-based beam-steered single-LED single-PD OWP architecture for 3D indoor localization. The core idea is to steer a single LED through $K$ predefined orientations while a PD at an unknown position---held at an arbitrary fixed orientation---collects the resulting $K$ RSS measurements. Forming the ratio of received powers between any two orientations cancels the unknown distance, transmitted power, and all receiver-orientation-dependent terms, yielding linear constraints on the transmitter-to-receiver direction vector; a closed-form eigenvector solution then provides the direction estimate. Because the ratio-based formulation eliminates all receiver-dependent factors, the direction-finding stage is mathematically independent of the receiver orientation $\mathbf{n}_r$, requiring no receiver pose knowledge or control.

This direction-finding capability constitutes the primary contribution of this work: once the transmitter-to-receiver direction is known, it directly enables beam steering toward the receiver to maximize the link signal-to-noise ratio (SNR) for a subsequent communication or ranging measurement. As an extension for full 3D positioning, a single additional cooperative measurement with the LED steered toward the estimated direction and the PD reoriented in the opposite direction from the estimated one, recovers the LED-to-PD distance, completing the 3D position estimate. Unlike prior single-LED approaches that require continuous PD rotation \cite{Liu2022, Wang2024, Shi2025}, the direction-finding stage requires zero receiver movement, and the entire beam-steering complexity resides in the infrastructure.

The main contributions are summarized as follows:
\begin{itemize}
    \item A novel single-LED single-PD 3D OWP framework that achieves full 3D localization without camera sensing or PD arrays, by shifting the spatial diversity from the receiver to the transmitter through $K$ steered beam orientations. We prove that all proposed direction-finding estimators, namely the generalized least squares (GLS), weighted least squares (WLS) and nonlinear least-squares (NLS) estimators, are mathematically independent of the receiver orientation $\mathbf{n}_r$, enabling direction finding without receiver pose knowledge.
    \item Direction error bound (DEB) and position error bound (PEB) analysis. We derive Fisher-information matrix (FIM) expressions, define the DEB as the Cramér--Rao lower bound (CRLB) for the angular direction error on $\mathbb{S}^2$, and characterize identifiability and FIM rank conditions as a function of $K$.
    \item Genetic algorithm (GA)-based orientation-set design. We cast the transmitter orientation selection as an offline GA optimization that minimizes the root-mean-square (RMS) DEB over a 3D testbed, yielding practical design rules for the choice of $K$ and the set of orientations.
    \item Closed-form linear direction estimators. Using a ratio-based linearization of the Lambertian power model, we develop a statistically efficient GLS direction estimator and a lightweight WLS approximation, both solvable via a single $3\times3$ eigendecomposition with microsecond-level latency.
    \item An iterative NLS direction estimator that fits the normalized Lambertian model on $\mathbb{S}^2$. This estimator is also $\mathbf{n}_r$-independent through power normalization, and achieves direction-finding accuracy closest to the DEB.
\end{itemize}

This paper is organized as follows.
Section~\ref{sec:SystemModel} presents the system model and the proposed 3D localization procedure.
Section~\ref{sec:PEB} derives the DEB and PEB.
Section~\ref{sec:optimization} optimizes the transmitter beam-orientation set.
Section~\ref{sec:NoLineal} develops the NLS direction estimator.
Section~\ref{sec:Lineal} introduces closed-form linear estimators (GLS and WLS), and establishes receiver-orientation independence.
Section~\ref{sec:results} reports the estimators comparison.
Section~\ref{sec:Conclusion} concludes the paper.

\paragraph*{Notation}
Bold lowercase letters denote column vectors (e.g., $\mathbf x$) and bold uppercase letters denote matrices (e.g., $\mathbf A$); scalars are in lightface. The Euclidean norm is $\lVert\cdot\rVert$, the inner product is $\mathbf a\!\cdot\!\mathbf b$, the trace is $\operatorname{tr}(\cdot)$, and the gradient with respect to $\mathbf{r}=[x,y,z]$ is $\nabla_{\mathbf r}=[\partial/\partial x,\partial/\partial y,\partial/\partial z]^{\mathsf T}$. The transpose is $(\cdot)^{\mathsf T}$, 
the identity matrix of dimension $n$ and the $n\times m$ zero matrix are $\mathbf{I}_n$ and $\mathbf{0}_{n\times m}$, respectively, and $\operatorname{diag}(\cdot)$ forms a diagonal matrix from its arguments. The smallest eigenvalue and a corresponding unit‐norm eigenvector of a symmetric matrix $\mathbf M$ are denoted by $\lambda_{\min}(\mathbf M)$ and $\mathbf v_{\min}(\mathbf M)$; the Loewner order $\succeq$ indicates positive semidefiniteness. $\mathbb R^n$ is the $n$-dimensional real space and $\mathbb S^{2}$ the unit sphere in $\mathbb R^{3}$. Random variables follow the convention $x\sim\mathcal N(\mu,\sigma^{2})$, with expectation $\mathbb E[\cdot]$ and variance $\operatorname{Var}[\cdot]$. A hat $(\hat{\cdot})$ denotes an estimate and an overbar $(\bar{\cdot})$ a sample mean.

\section{System Model and Proposed \\3D Localization Method}
\label{sec:SystemModel}
\subsection{System Model}

\begin{figure}[t]
    \centering
    \includegraphics[width=\linewidth]{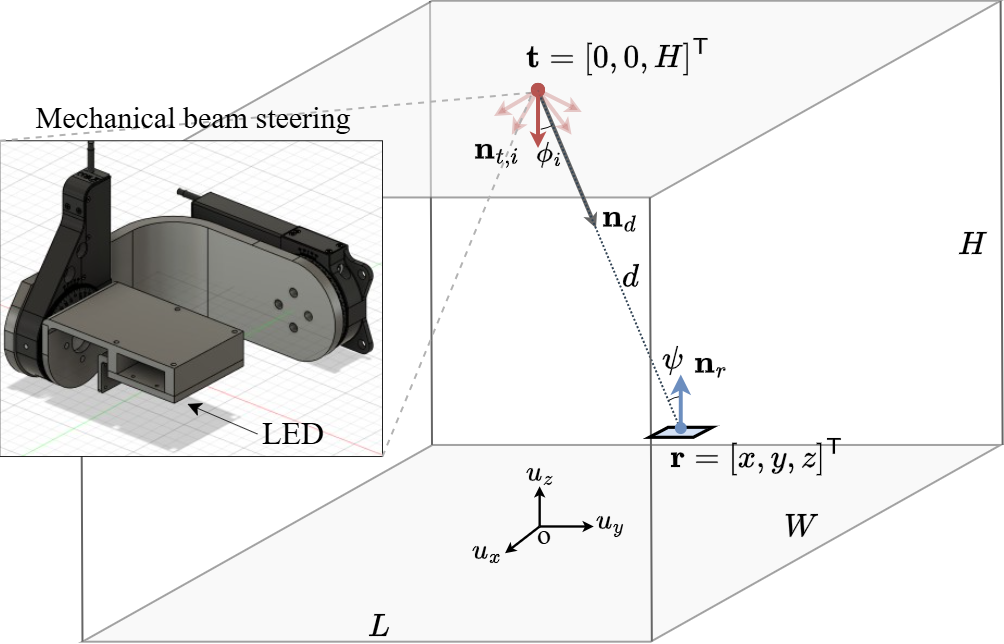}
    \caption{Single beam-steered LED / single-PD OWP geometry. A beam-steered LED at $\mathbf t=[0,0,H]^{\mathsf T}$ (e.g., mechanical steering) points along $\{\mathbf n_{t,i}\}_{i=1}^{K}$; the PD is at $\mathbf r=[x,y,z]^{\mathsf T}$.}
    \label{fig:system}
\end{figure}

Figure~\ref{fig:system} illustrates the single-LED single-PD OWP setup analyzed in this work. A LED transmitter is mounted on the ceiling at a known position \(\mathbf{t}=[0,0,H]^{\mathsf T}\) and can steer its optical axis along a unit vector \(\mathbf{n}_{t,i}\!\in\!\mathbb{S}^{2}\) for the \(i\)-th orientation, \(i=1,\dots,K\). The receiver is a single PD with normal vector \(\mathbf{n}_{r}\) $\in\mathbb{S}^2$ located at an unknown position \(\mathbf{r}=[x,y,z]^{\mathsf T}\). For concreteness, we set $\mathbf{n}_r=[0,0,1]^{\mathsf T}$ in the simulations. As shown in Section~\ref{subsec:nr_independence}, all proposed direction estimators are independent of $\mathbf{n}_r$; therefore, the direction-finding results hold for any receiver orientation. Let \(\mathbf{d}=\mathbf{r}-\mathbf{t}\) denote the transmitter--receiver displacement vector, \(d=\lVert \mathbf{d}\rVert\) the separation distance, and \(\mathbf{n}_{d}= \mathbf{d}/d\) the corresponding unit direction vector.

For the \(i\)-th LED orientation, the RSS at the PD is modeled as \cite{Kahn1997}:
\begin{equation}
S_{r,i} = R_pP_{t}\,h_{\text{LOS},i} + w_i,
\label{eq:Sr_i}
\end{equation}
where \(R_p\) is the PD responsivity,
\(P_{t}\) is the transmitted radiant power,
and \(w_i \sim \mathcal{N}(0,\sigma_{w}^{2})\) models a zero-mean additive white Gaussian noise (AWGN) term arising primarily from shot and Johnson noise, and assumed independent and identically distributed (i.i.d.) across orientations and time (homoscedastic), and \(h_{\text{LOS},i}\) denotes the LOS channel gain, which for a Lambertian source is given by \cite{Kahn1997}:
\begin{equation}
h_{\text{LOS},i} =
\begin{cases}
\dfrac{(m+1)A_{\text{det}}}{2\pi d^{2}}
      \cos^{m}(\phi_{i})\cos(\psi), & \psi \leq \Psi_{\text{FOV}}, \\[6pt]
0, & \text{otherwise},
\end{cases}
\label{eq:hLOS}
\end{equation}
where \(m\) is the Lambertian order of the LED, related to the semi-angle at half power \(\Phi_{1/2}\) by \(m = -\ln2/\ln(\cos(\Phi_{1/2}))\), \(A_{\text{det}}\) the PD effective area, \(\Psi_{\text{FOV}}\) the receiver FOV, and the irradiance/incidence cosines geometrically given by:
\begin{equation}
\cos\phi_{i}= \dfrac{\mathbf{n}_{t,i}\!\cdot\!\mathbf{d}}{d},
\qquad
\cos\psi    = \dfrac{\mathbf{n}_{r}\!\cdot(-\mathbf{d})}{d}.
\label{eq:cosines}
\end{equation}

Dividing \eqref{eq:Sr_i} by \(R_p\) yields the received optical power at a given time instant:
\begin{equation}
P_{r,i} = P_{t}\,h_{\text{LOS},i} + n_i,
\label{eq:Pr_i}
\end{equation}
where \(n_i \sim \mathcal{N}(0,\sigma^{2})\) is zero-mean AWGN, remains i.i.d. and homoscedastic, with variance \(\sigma^{2}=(\sigma_{w}^{2}/R_p^{2})\).

At every transmitter orientation \(i\), the receiver acquires \(N\) independent samples. The sample mean of the received power is:
\begin{equation}
\bar P_{r,i}=\frac{1}{N}\sum_{k=1}^{N} P_{r,i,k},
\label{eq:sampleMean}
\end{equation}
where \(P_{r,i,k}\) denotes the \(k\)-th instantaneous received power sample at orientation \(i\), corresponding to the \(k\)-th realization of the random variable \(P_{r,i}\) in \eqref{eq:Pr_i}. Under i.i.d. AWGN, \(\bar P_{r,i}\) is the minimum-variance unbiased (MVU) estimator of the received power.

The corresponding SNR, given the receiver position and a specific transmitter orientation \(i\), denoted $\mathrm{SNR}_{r,i}$, as well as the average SNR for a given receiver position, noted $\mathrm{SNR}_{r}$, and for the entire testbed, noted $\mathrm{SNR}$, are respectively defined as:
\begin{subequations}\label{eq:SNR}
\begin{align}
\mathrm{SNR}_{r,i} &= \frac{(R_pP_{t}\,h_{\text{LOS},i})^2}{\sigma_{w}^{2}}, \label{eq:SNRa} \\
\mathrm{SNR}_{r} &= \frac{1}{K}\sum_{i=1}^{K}\mathrm{SNR}_{r,i}, \label{eq:SNRb} \\
\mathrm{SNR} &= \frac{1}{|\mathcal{R}|}\sum_{r\in\mathcal{R}}{\mathrm{SNR}}_{r}, \label{eq:SNRc}
\end{align}
\end{subequations}
where \(\mathcal{R}\) denotes the set of receiver testbed positions and $|\mathcal R|$ their cardinality.

\subsection{General 3D Single-LED Positioning Procedure}

The proposed localization process comprises two consecutive stages:

\subsubsection{Direction Finding}

The first stage estimates the direction vector \(\mathbf{n}_{d}\) by steering the LED through the \(K\) predefined orientations \(\{\mathbf{n}_{t,i}\}_{i=1}^{K}\) and processing the power means \(\{\bar P_{r,i}\}_{i=1}^{K}\). Section~\ref{sec:NoLineal} (NLS estimator) and Section~\ref{sec:Lineal} (linear GLS/WLS estimators) detail statistically grounded solutions.

\subsubsection{Distance Recovery}

After direction finding, the LED is steered to the estimated direction $\hat{\mathbf{n}}_{d}$ so that its beam is directed toward the receiver, and the PD is reoriented to $\mathbf{n}_r = -\hat{\mathbf{n}}_d$ so that it faces the transmitter. Under this beam-steered configuration and using \eqref{eq:hLOS}, \eqref{eq:cosines}, and \eqref{eq:Pr_i}, the received optical power at a given time instant for orientation $(K+1)$ obeys:
\begin{equation}
P_{r,K+1}= \frac{C}{d^{2}}+n_{K+1},
\label{eq:beamSteeredMean}
\end{equation}
where \(C\) is an optical constant calibrated once per hardware set-up:
\begin{equation}
C = \frac{P_{t}(m+1)A_{\text{det}}}{2\pi},
\label{eq:constant_C}
\end{equation} 
and $n_{K+1}\sim\mathcal{N}(0,\sigma^{2})$ is an i.i.d.\ noise realization at orientation $(K+1)$, with the same distribution as in \eqref{eq:Pr_i}.

Collecting \(N\) independent samples \(\{P_{r,K+1,j}\}_{j=1}^{N}\) and using~\eqref{eq:sampleMean}, the sample mean satisfies \(\bar P_{r,K+1}\sim\mathcal{N}\bigl(C/d^{2},\sigma^{2}/N\bigr)\). Substituting \(\bar P_{r,K+1}\) into~\eqref{eq:beamSteeredMean} provides the distance estimate:
\begin{equation}
\hat d = \sqrt{\frac{C}{\bar P_{r,K+1}}},
\label{eq:dhat}
\end{equation}
and, finally, the 3D estimated position of the receiver, denoted as \(\hat{\mathbf{r}}\), is $\hat{\mathbf{r}} = \mathbf{t} + \hat d \,\hat{\mathbf{n}}_{d}$.

This cooperative alignment maximizes both $\cos\phi\approx 1$ and $\cos\psi\approx 1$, thereby maximizing the received power and the SNR of the ranging measurement. The PD reorientation is performed once, after the $K$ direction-finding measurements during which the PD remains at an arbitrary fixed pose. This contrasts with prior single-LED approaches that require continuous PD rotation \cite{Liu2022,Wang2024,Shi2025}.

\section{Direction Error Bound and Position Error Bound}
\label{sec:PEB}

This section derives the CRLB for direction estimation from the $K$ steered-orientation measurements. We term this limit the DEB. As an extension, we also define the PEB, which additionally accounts for the distance-recovery measurement and serves as a benchmark for the full 3D pipeline.

\subsection{Direction-Finding Observation Model}
At each transmitter orientation $i=1,\ldots,K$, the sample-mean received power~\eqref{eq:sampleMean} is distributed as:
\begin{equation}
\bar P_{r,i} \sim 
  \mathcal{N}\!\bigl(\mu_{i}(\mathbf r),\,\sigma^{2}/N\bigr),
\quad i=1,\dots,K.
\end{equation}
From~\eqref{eq:Pr_i}--\eqref{eq:cosines}, the noise-free mean power factorizes as:
\begin{equation}
\mu_i = \eta\,Q_i^{\,m},
\label{eq:mu_factorized}
\end{equation}
where:
\begin{equation}
Q_i = \mathbf{n}_{t,i}\cdot\mathbf{n}_d = \cos\phi_i
\label{eq:Q}
\end{equation}
is the direction cosine between the $i$-th LED orientation and the true transmitter-to-receiver direction $\mathbf{n}_d$, and:
\begin{equation}
\eta = \frac{C\,\cos\psi}{d^2}
\label{eq:eta_nuisance}
\end{equation}
is a positive scalar common to all $K$ orientations.  The constant $\eta$ absorbs the unknown distance~$d$ and the receiver orientation $\mathbf{n}_r$ (through $\cos\psi$).  Since the direction-finding stage aims to recover $\mathbf{n}_d$ alone, $\eta$ is a nuisance parameter that must be accounted for in the bound derivation.

\subsection{Fisher Information for Direction Estimation}
\label{subsec:FIM}
We parameterize the direction in nadir-referenced spherical coordinates:
\begin{equation}
\mathbf{n}_d = \bigl[\sin\theta_d\cos\phi_d,\;\sin\theta_d\sin\phi_d,\;-\cos\theta_d\bigr]^{\mathsf T},
\label{eq:nd_spherical}
\end{equation}
where $\theta_d\in[0,\pi)$ is the polar angle from $-\mathbf{u}_z$ and $\phi_d\in[0,2\pi)$ is the azimuth.  Let $\mathbf{u}_\theta = \partial\mathbf{n}_d/\partial\theta_d$ and $\mathbf{u}_\phi = \partial\mathbf{n}_d/\partial\phi_d$ denote the tangent vectors to $\mathbb{S}^2$ at $\mathbf{n}_d$.

We adopt $\alpha\triangleq\ln\eta$ as the nuisance coordinate (the log-reparameterization ensures positivity and improves numerical conditioning) and form the extended parameter vector $\boldsymbol{\xi}=[\theta_d,\,\phi_d,\,\alpha]^{\mathsf T}\in\mathbb{R}^3$.  Under i.i.d.\ AWGN with variance $\sigma^2/N$, the Slepian--Bangs formula~\cite{Kay1993} yields the $3\times3$ FIM:
\begin{equation}
\mathcal{I}_{\mathrm{DF}}(\boldsymbol\xi)
= \frac{N\,\eta^2}{\sigma^2}
  \sum_{i=1}^{K}
  \tilde{\mathbf g}_i\,\tilde{\mathbf g}_i^{\mathsf T},
\label{eq:FIM_DEB}
\end{equation}
where the normalized gradient vector for the $i$-th orientation is:
\begin{equation}
\tilde{\mathbf g}_i = \bigl[m\,Q_i^{m-1}(\mathbf{n}_{t,i}\!\cdot\!\mathbf{u}_\theta),\;\;m\,Q_i^{m-1}(\mathbf{n}_{t,i}\!\cdot\!\mathbf{u}_\phi),\;\;Q_i^m\bigr]^{\mathsf T}.
\label{eq:grad_xi}
\end{equation}
The first two components capture the sensitivity of $\mu_i$ to angular perturbations of $\mathbf{n}_d$, while the third component captures sensitivity to the amplitude $\eta$.

\subsection{Direction Error Bound}
\label{subsec:DEB}
Since $\alpha$ is not of interest, it is profiled out via the Schur complement of the $(3,3)$ block of $\mathcal{I}_{\mathrm{DF}}$~\cite{Kay1993}.  The resulting CRLB for the angular parameters $(\theta_d,\phi_d)$ is the upper-left $2\times2$ block of the inverse FIM:
\begin{equation}
\mathbf{C}_{\mathrm{ang}} = \bigl[\mathcal{I}_{\mathrm{DF}}^{-1}(\boldsymbol\xi)\bigr]_{1:2,\,1:2}.
\label{eq:C_ang}
\end{equation}
This matrix provides a tight lower bound on the covariance of any unbiased estimator of $(\theta_d,\phi_d)$, irrespective of how the nuisance parameter $\eta$ is handled (estimated jointly, profiled, or known).

To express the bound in terms of the Cartesian direction error $\lVert\hat{\mathbf{n}}_d - \mathbf{n}_d\rVert$, we project through the Jacobian $\mathbf{J}_{\mathrm{sph}}=[\mathbf{u}_\theta,\,\mathbf{u}_\phi]\in\mathbb{R}^{3\times2}$:
\begin{equation}
\mathrm{DEB}(\mathbf{r})
= \sqrt{\operatorname{tr}\!\bigl\{\mathbf{J}_{\mathrm{sph}}\,\mathbf{C}_{\mathrm{ang}}\,\mathbf{J}_{\mathrm{sph}}^{\mathsf T}\bigr\}}.
\label{eq:DEB}
\end{equation}
For small errors, $\mathrm{DEB}\approx$ angular RMS error (RMSE) in radians. The DEB depends on the receiver position~$\mathbf{r}$ (through $\eta$ and $\{Q_i\}$), on the orientation set $\{\mathbf{n}_{t,i}\}$, and on the system parameters $(N,\sigma^2,m,P_t,A_{\mathrm{det}})$.  Crucially, it uses only the $K$ direction-finding measurements and is therefore the appropriate benchmark for direction estimators.

\subsection{Position Error Bound}
\label{subsec:PEB}
When the full two-stage pipeline is evaluated (direction finding followed by distance recovery), an additional measurement $\bar P_{r,K+1}\sim\mathcal{N}(C/d^2,\,\sigma^2/N)$ becomes available.  Including this $(K{+}1)$-th observation in the FIM yields the Cartesian-parameterized $3\times3$ matrix:
\begin{equation}
\mathcal{I}(\mathbf r)
= \frac{N}{\sigma^{2}}
  \sum_{i=1}^{K+1}
  \bigl[\nabla_{\mathbf r}\mu_i(\mathbf r)\bigr]\,
  \bigl[\nabla_{\mathbf r}\mu_i(\mathbf r)\bigr]^{\!\mathsf T},
\label{eq:FIM_general}
\end{equation}
where $\nabla_{\mathbf r}\mu_i$ for $i=1,\ldots,K$ is given by:
\begin{multline}
    \nabla_{\mathbf r}\mu_i(\mathbf r)
    = \frac{C}{d^{3}}
      \Bigl[
        m\,\cos^{m-1}\!\phi_i\;\cos\psi\;\mathbf n_{t,i}
        -
        \cos^{m}\!\phi_i\;\mathbf n_{r}
    \\[-3pt]
        -\;
        (m+3)\,\cos^{m}\!\phi_i\;\cos\psi\;\mathbf n_{d}
      \Bigr],
    \label{eq:grad_mu_closed}
\end{multline}
and the distance-recovery gradient is $\nabla_{\mathbf r}\mu_{K+1} = -2C\,\mathbf n_d/d^3$.  The PEB is defined as:
\begin{equation}
\mathrm{PEB}(\mathbf r)
= \sqrt{\operatorname{tr}\!\bigl\{\mathcal{I}^{-1}(\mathbf r)\bigr\}}.
\label{eq:PEB}
\end{equation}
Since the PEB assumes perfect beam alignment in the $(K{+}1)$-th measurement, it is an optimistic lower bound that no practical two-stage estimator can attain~\cite{Kay1993}.  In this work, the DEB~\eqref{eq:DEB} is adopted as the primary metric and GA objective; the PEB is reported in Section~\ref{sec:results} for completeness.

\subsection{Number of LED Orientations}
\label{subsec:Kchoice}

Since the DEB is the primary metric adopted in this work, the identifiability conditions are stated in terms of the direction-finding FIM $\mathcal{I}_{\mathrm{DF}}(\boldsymbol\xi)$ in~\eqref{eq:FIM_DEB}, which is a sum of $K$ rank-one outer products $\tilde{\mathbf g}_i\tilde{\mathbf g}_i^{\mathsf T}$.

\subsubsection{Underdetermined Regime ($K\le2$)}
When $K\le2$, at most two gradient vectors $\{\tilde{\mathbf g}_i\}$ are available, so $\operatorname{rank}(\mathcal{I}_{\mathrm{DF}})\le 2<3$. The FIM is singular, $\mathcal{I}_{\mathrm{DF}}^{-1}$ does not exist, and the DEB~\eqref{eq:DEB} diverges \cite{Zhuang2018}.

\subsubsection{Just-Determined Regime ($K=3$)}
With three orientations whose gradient vectors $\{\tilde{\mathbf g}_1,\tilde{\mathbf g}_2,\tilde{\mathbf g}_3\}$ are linearly independent---which requires non-coplanar LED orientations---the FIM reaches full rank and $\mathcal{I}_{\mathrm{DF}}^{-1}$ exists, yielding a finite DEB. However, the absence of redundancy makes the bound highly sensitive to small perturbations in orientation geometry \cite{Chen2021}.

\subsubsection{Overdetermined Regime ($K\ge4$)}
For four or more orientations, additional rank-one contributions improve the numerical conditioning of $\mathcal{I}_{\mathrm{DF}}$ and lower the DEB, since each extra orientation provides independent angular information beyond the minimal three \cite{Ma2024,Wang2024b}. The same conclusions extend to the Cartesian FIM $\mathcal{I}(\mathbf r)$ in~\eqref{eq:FIM_general} and hence to the PEB, with the additional distance-recovery measurement further augmenting the information.

\section{Optimization of the Orientation Set}
\label{sec:optimization}

The DEB in~\eqref{eq:DEB} depends explicitly on the transmitter orientation set $\{\mathbf{n}_{t,i}\}_{i=1}^K$ through the direction-finding FIM $\mathcal{I}_{\mathrm{DF}}(\boldsymbol\xi)$ in~\eqref{eq:FIM_DEB}. We define the objective function as the RMSE of the DEB evaluated over all receiver positions in the 3D testbed specified in Table~\ref{tab:ga_params}:
\begin{equation}
f(\mathbf{x}) = \sqrt{\frac{1}{|\mathcal{R}|}\sum_{\mathbf{r}\in\mathcal{R}}\mathrm{DEB}^2(\mathbf{r};\,\mathbf{x})},
\label{eq:GA_objective}
\end{equation}
where $\mathbf{x}=[\theta_1,\varphi_1,\ldots,\theta_K,\varphi_K]^{\mathsf T}$ encodes the orientation set.
Selecting the orientation set that minimizes this objective is essential for direction-finding accuracy. However, \(\mathrm{DEB}(\{\mathbf n_{t,i}\})\) is nonlinear in the orientation parameters, involves multiple coupled terms (e.g., powers of cosines and distance-dependent factors), and exhibits nonsmooth behavior due to the receiver FOV. Consequently, closed-form optimization is not available and purely local methods are prone to suboptimal solutions. To robustly explore the parameter space and approach a globally optimal orientation set for a given \(K\), we employ a GA \cite{Haupt2004}.

\begin{table}[tbh]
\centering
\caption{Parameters used for GA optimization.}
\label{tab:ga_params}
\begin{tabular}{|l|l|}
\hline
\multicolumn{2}{|c|}{\textbf{System Parameters}} \\ \hline
\textbf{Parameter}                  & \textbf{Value}                         \\ \hline
Transmitter coordinates $(\mathbf{t})$ & \([0,0,2]^{\mathsf T}\,\mathrm{m}\)              \\ \hline
Transmitted optical power $(P_t)$   & \(0.405\,\mathrm{W}\)            \\ \hline
Semi-angle at half power  $(\Phi_{1/2})$ & \(\{30^\circ,45^\circ,60^\circ\}\)                \\ \hline
Lambertian order $(m)$              & \(-\ln2/\ln(\cos(\Phi_{1/2}))\)    \\ \hline
Photodiode effective area $(A_{\mathrm{det}})$        & \(4.8\times5.5\,\mathrm{mm}^2\) \\ \hline
Receiver FOV $(\Psi_{\mathrm{FOV}})$        & \(85^\circ\)       \\ \hline
Electrical noise variance $(\sigma_w^2)$ & \(1.19\times10^{-14}\,\mathrm{A}^2\) \\ \hline
PD responsivity $(R_{p})$ & \(0.63\,\mathrm{A/W}\)          \\ \hline
Noise variance $(\sigma^2)$         & \(3\times10^{-14}\,\mathrm{W}^2\) \\ \hline
Samples per orientation $(N)$       & \(1000\)                           \\ \hline
\multicolumn{2}{|c|}{\textbf{Genetic Algorithm}} \\ \hline
\textbf{Parameter}                  & \textbf{Value}                         \\ \hline
Population size $(P)$                    & 300                                    \\ \hline
Maximum generations $(G_{\mathrm{max}})$     & 200                                    \\ \hline
Crossover probability $(p_c)$       & 0.8                                    \\ \hline
Search range $(\Omega)$             & \(\theta_{i}\in[0,80^\circ],\;\varphi_{i}\in[0,360^\circ]\) \\ \hline
Objective function $(f(\mathbf x))$   & RMSE of the DEB~\eqref{eq:GA_objective}                        \\ \hline
Decision variables $(\mathbf x)$    & \([\theta_{1},\varphi_{1},\dots,\theta_{K},\varphi_{K}]^{\mathsf T}\) \\ \hline
\multicolumn{2}{|c|}{\textbf{3D Testbed}} \\ \hline
\textbf{Parameter}                  & \textbf{Value}                         \\ \hline
Room dimensions \((L\times W \times H)\) & \(3 \times 3 \times 2\,\mathrm{m^3}\)  \\ \hline
Range in \(x,y\)                    & \([-1.5,\,1.5]\,\mathrm{m}\)           \\ \hline
Range in \(z\)                      & \([0,\,1.2]\,\mathrm{m}\)              \\ \hline
Grid step                           & \(0.2\,\mathrm{m}\)                    \\ \hline
Total number of points tested $|\mathcal{R}|$  & \(1792\)                               \\ \hline
\end{tabular}
\end{table}

Within this GA framework, each individual encodes \(K\) steering directions via tilt–azimuth pairs \(\{(\theta_i,\varphi_i)\}_{i=1}^{K}\), yielding a \(2K\)-dimensional genotype \(\mathbf x = [\theta_1,\varphi_1,\dots,\theta_K,\varphi_K]^{\mathsf T}\).

We adopt a nadir‐referenced spherical parameterisation where $\mathbf u_x=[1,0,0]^{\mathsf T}$, $\mathbf u_y=[0,1,0]^{\mathsf T}$, $\mathbf u_z=[0,0,1]^{\mathsf T}$ are the standard Cartesian basis vectors, \(\theta_i\) is the tilt from \(-\mathbf u_z\) and \(\varphi_i\) is the azimuth in the \(\mathbf u_x\)–\( \mathbf u_y\) plane measured counterclockwise from \(+\mathbf u_x\), with domains \(\theta_i\in[0,\theta_{\max}]\) and \(\varphi_i\in[0^\circ,360^\circ]\). The mapping to unit‐norm orientation vectors in the reference frame \((\mathbf u_x,\mathbf u_y,\mathbf u_z)\) is:
\begin{equation}
\mathbf n_{t,i}
= \bigl[\sin\theta_i\cos\varphi_i,\;\sin\theta_i\sin\varphi_i,\;-\,\cos\theta_i\bigr]^{\mathsf T}.
\end{equation}
Spherical coordinates are preferred to Cartesian because they describe unit‐norm directions with two angles, avoiding the nonlinear unit‐norm constraint and providing bounded domains that simplify GA search.
Standard genetic operators—tournament selection, crossover with probability \(p_c\), and adaptive feasible mutation—are applied for at most \(G_{\max}\) generations, 
with early termination when the average relative improvement of the best fitness over \(G_{\max}\) generations falls below \(10^{-6}\).
The implementation is in MATLAB; optimization parameters are summarized in Table~\ref{tab:ga_params}.

We emphasize that the GA optimization is an offline system-design problem, not an online estimation problem. The $2K$ decision variables (tilt--azimuth pairs) define the orientation set; for each candidate set, the DEB (or PEB) is evaluated analytically over the entire 3D testbed using the closed-form FIM in~\eqref{eq:FIM_general}. No measurements are involved in this optimization---only theoretical performance bounds. The orientation set is optimized to minimize the RMSE of the DEB averaged over all $|\mathcal{R}|=1{,}792$ receiver positions (Table~\ref{tab:ga_params}). The position $\mathbf{r}$ is therefore not fixed but sweeps the spatial grid during fitness evaluation; the resulting orientation set is position-independent and can be deployed universally. Once fixed, the online direction-finding stage uses $K$ power measurements to recover the two-degrees-of-freedom direction on $\mathbb{S}^2$, which is overdetermined for $K\ge3$.

\begin{table*}[ht]
\centering
\caption{DEB-optimized orientations \((\theta_i,\varphi_i)\) in degrees \([^\circ]\) for various \(K\) in a \(3\times3\times2\,\mathrm{m^3}\) room, with $\Phi_{1/2}=45^\circ$ and the achieved RMS-DEB.}
\label{tab:optimal_orientations_2col}
\resizebox{\linewidth}{!}{\begin{tabular}{|c|c|*{9}{c|}}
\hline
\(K\) & \textbf{RMS-DEB}  & \(i=1\)                & \(i=2\)                & \(i=3\)                & \(i=4\)                & \(i=5\)                & \(i=6\)                & \(i=7\)                & \(i=8\)                & \(i=9\)                         \\ 
\hline
\(3\) & $1.30^\circ$  & \((17.5,\,203.7)\)   & \((17.2,\,332.2)\)   & \((18.7,\,88.8)\)    &   -                    &   -                    &   -                    &   -                    &   -                    &   -                         \\ \hline
\(4\) & $0.75^\circ$  & \((29.9,\,315.0)\)   & \((29.9,\,135.0)\)   & \((29.9,\,45.0)\)    & \((29.9,\,225.0)\)     &   -                    &   -                    &   -                    &   -                    &   -                         \\ \hline
\(5\) & $0.52^\circ$  & \((0.1,\,74.2)\)     & \((65.7,\,269.8)\)   & \((65.7,\,179.9)\)   & \((65.9,\,359.8)\)     & \((65.9,\,89.9)\)      &   -                    &   -                    &   -                    &   -                         \\ \hline
\(6\) & $0.47^\circ$  & \((67.6,\,253.4)\)   & \((66.6,\,321.1)\)   & \((70.0,\,176.9)\)   & \((0.1,\,274.3)\)      & \((68.9,\,98.4)\)      & \((66.8,\,24.9)\)      &   -                    &   -                    &   -                         \\ \hline
\(7\) & $0.41^\circ$  & \((65.6,\,353.3)\)   & \((2.5,\,10.1)\)     & \((66.2,\,273.6)\)   & \((64.6,\,196.5)\)     & \((62.5,\,130.8)\)     & \((2.6,\,191.7)\)      & \((63.8,\,68.1)\)      &   -                    &   -                         \\ \hline
\(8\) & $0.39^\circ$  & \((67.6,\,67.1)\)    & \((66.6,\,247.3)\)   & \((2.8,\,39.7)\)     & \((3.2,\,227.0)\)      & \((64.7,\,2.8)\)       & \((66.1,\,179.7)\)     & \((66.9,\,298.2)\)     & \((65.4,\,114.9)\)     &   -                         \\ \hline
\(9\) & $0.36^\circ$  & \((66.1,\,165.4)\)   & \((8.7,\,267.1)\)    & \((66.8,\,273.8)\)   & \((62.4,\,219.4)\)     & \((64.2,\,21.6)\)      & \((67.2,\,91.0)\)      & \((13.9,\,105.1)\)     & \((4.5,\,303.1)\)      & \((62.0,\,330.3)\)          \\
\hline
\end{tabular}}
\end{table*}

Table~\ref{tab:optimal_orientations_2col} lists the DEB-optimal \(K\)-orientation sets for the system parameters in Table~\ref{tab:ga_params} when $\Phi_{1/2} = 45^{\circ}$, together with the achieved RMS-DEB. Each set is unique and no set is a subset of another. The RMS-DEB decreases monotonically from $1.30^\circ$ at $K=3$ to $0.36^\circ$ at $K=9$, showing the benefit of additional orientations for direction-finding accuracy. At the selected operating point $K=5$, the DEB-optimized set achieves an RMS-DEB of $0.52^\circ$.

\begin{figure}[tb]
    \centering
    \includegraphics[width=\linewidth]{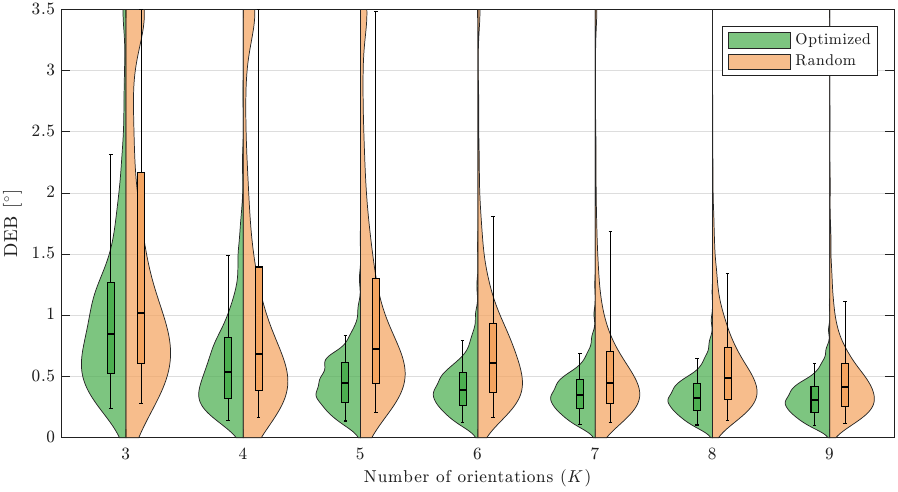}
    \caption{Distribution of the DEB [$^\circ$] across all receiver locations in the $3\times3\times2\,\mathrm{m^3}$ testbed for $K\in\{3,\ldots,9\}$, with $\Phi_{1/2} = 45^{\circ}$. Green: GA-optimized orientation set; orange: 50 independently drawn random orientation sets per~$K$. Box-and-whisker plots summarize the sample distribution; shaded profiles depict the estimated PDF. The GA-optimized sets consistently achieve lower DEB and tighter distributions across all~$K$.}
    \label{fig:Optimized_vs_Random}
\end{figure}

In Fig.~\ref{fig:Optimized_vs_Random}, for each \(K\) we present box-and-whisker plots of the DEB computed over all receiver locations in the \(3\times3\times2\,\mathrm{m}^3\) testbed; optimized sets are shown in green, randomly selected sets in orange, and the shaded profiles depict the corresponding probability density function (PDF) estimates. The GA consistently outperforms random orientations: across \(K\), the optimized sets reduce the median DEB by \(49\%\)–\(72\%\) and exhibit smaller interquartile ranges. These gains are reflected in the distributions, with the central tendency improving monotonically as \(K\) increases.

\begin{figure}[tb]
  \centering
  \includegraphics[width=\columnwidth]{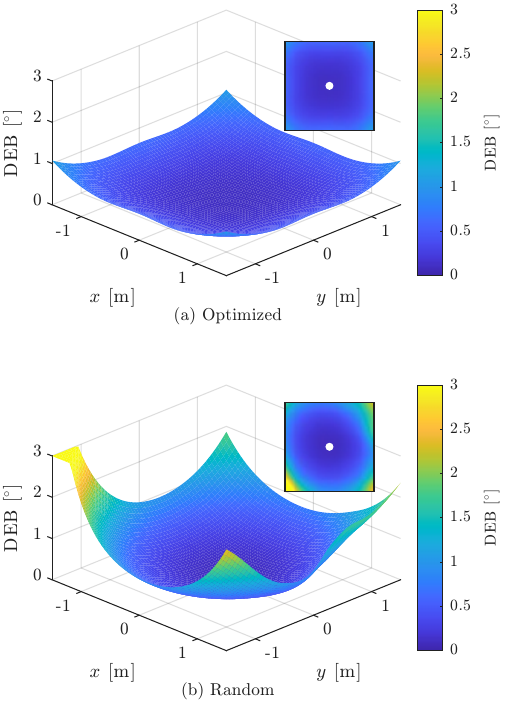}
  \caption{Spatial distribution of the DEB across the $3\times3\,\mathrm{m^2}$ floor at $z=0.8$\,m for $K=5$ orientations, and with $\Phi_{1/2} = 45^{\circ}$. (a)~GA-optimized set; (b)~a single random set with tilt--azimuth pairs drawn uniformly from the GA search range. The color scale is shared and clipped at $3^\circ$.}
  \label{fig:optimo_vs_random_3D}
\end{figure}

Figure~\ref{fig:optimo_vs_random_3D} presents spatial heat maps of the DEB across a \(3\times3\,\mathrm{m^2}\) floor at \(z=0.8\) m for \(K=5\) orientations with the transmitter at the room centre.  In the optimal configuration, shown in Fig.~\ref{fig:optimo_vs_random_3D}(a), the highest errors occur uniformly along the boundaries and especially at the corners, analogous to multi‐transmitter layouts \cite{acuna2025}, but remain spatially consistent.  By contrast, the random configuration in Fig.~\ref{fig:optimo_vs_random_3D}(b) exhibits both larger and more irregular error patterns.  

\begin{figure}[tb]
    \centering
    \includegraphics[width=\linewidth]{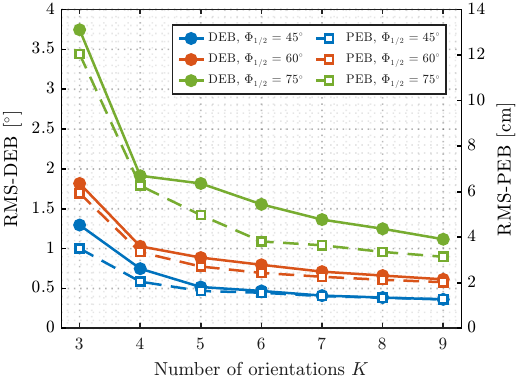}
    \caption{RMS-DEB (solid, left axis) and RMS-PEB (dashed, right axis) versus the number of orientations~$K$ for the DEB-optimized sets and three LED semi-angles at half power $\Phi_{1/2}\in\{45^\circ,60^\circ,75^\circ\}$. Values are computed over all receiver positions in the $3\times3\times2~\mathrm{m}^3$ testbed. Both bounds decrease monotonically with~$K$; the operating point $K=5$ offers a favourable accuracy--latency trade-off for both direction finding and 3D positioning.}
    \label{fig:PEB_vs_K}
\end{figure}

We now turn from optimization to performance characterization. Figure~\ref{fig:PEB_vs_K} plots the RMS-DEB (solid, left axis) and RMS-PEB (dashed, right axis) versus~$K$ for the DEB-optimized orientation sets under three semi-angles at half power $\Phi_{1/2}\in\{45^\circ,60^\circ,75^\circ\}$. Both bounds decrease monotonically with~$K$, confirming that additional orientations improve direction-finding and positioning accuracy simultaneously. The direction-finding bound reaches sub-degree RMS-DEB at $K\ge5$ for $\Phi_{1/2}=45^\circ$, while the corresponding RMS-PEB falls below~$2$\,cm. Wider beams ($\Phi_{1/2}=60^\circ$--$75^\circ$) incur higher bounds due to the reduced Lambertian directivity, yet the trends remain consistent. Assuming a fixed per-orientation acquisition time, latency scales linearly with~$K$; under this accuracy--latency trade-off, $K=5$ offers a favourable compromise for both bounds. $\Phi_{1/2}=45^\circ$ achieves competitive performance starting at $K=3$ and, given its widespread use in commercial LEDs, is adopted for the subsequent simulations.

Figure~\ref{fig:PEB_vs_noise} presents the RMS-DEB (solid, left axis) and RMS-PEB (dashed, right axis) versus SNR for the DEB-optimized orientation sets with \(\Phi_{1/2}=45^\circ\). The bold curves correspond to the recommended operating point \(K=5\); the shaded bands span the range \(K\in\{3,\ldots,9\}\). Both bounds decrease monotonically with SNR and exhibit nearly identical slopes of \(-1/2\) on the log--log axes, confirming the theoretical \(1/\sqrt{\mathrm{SNR}}\) scaling of the CRLB under AWGN~\cite{Kay1993}. At a representative SNR of~\(14\)\,dB, the \(K=5\) set achieves an RMS-DEB below~\(0.3^\circ\) and an RMS-PEB below~\(1\)\,cm. The band width at any fixed SNR quantifies the gain from additional orientations; it narrows at high SNR, indicating diminishing returns from increasing~\(K\) in low-noise regimes.

\section{Nonlinear Least-Squares Direction Estimator}
\label{sec:NoLineal}

This section develops an iterative NLS direction estimator that operates directly on the Lambertian power model without the linearization employed by GLS/WLS (Section~\ref{sec:Lineal}). The key idea is to normalize the received powers so as to eliminate the common multiplicative factor that depends on the unknown distance~$d$ and receiver orientation~$\mathbf{n}_r$, thereby rendering the estimator $\mathbf{n}_r$-independent.

\subsection{NLS Cost Function and Direction Estimate}
As established in Section~\ref{subsec:FIM}, the noise-free received power factorizes as $\mu_i = \eta\,Q_i^m$~\eqref{eq:mu_factorized}, where $Q_i = \mathbf{n}_{t,i}\cdot\mathbf{n}_d$~\eqref{eq:Q} and $\eta = C\cos\psi/d^2$~\eqref{eq:eta_nuisance} is a common positive scalar that absorbs the unknown distance and receiver orientation.

Let $\hat\mu_i = \bar P_{r,i}$ denote the sample-mean power at orientation~$i$ (cf.~\eqref{eq:sampleMean}). We define the normalized power targets as:
\begin{equation}
p_i \triangleq \frac{\hat\mu_i}{\max_{j=1,\ldots,K}\hat\mu_j}, \quad i=1,\ldots,K.
\label{eq:p_target}
\end{equation}
In the noiseless regime, substituting~\eqref{eq:mu_factorized} yields:
\begin{equation}
p_i = \frac{\eta\,Q_i^m}{\eta\,Q_{\max}^m} = \frac{Q_i^m}{Q_{\max}^m},
\label{eq:p_target_noiseless}
\end{equation}
where $Q_{\max}=\max_j Q_j$. The common factor~$\eta$ cancels exactly, so the normalized targets $p_i$ depend only on the LED orientations $\{\mathbf{n}_{t,i}\}$ and the direction $\mathbf{n}_d$, not on $\mathbf{n}_r$, $d$, $P_t$, or $A_{\mathrm{det}}$.

We model the normalized targets as $p_i \approx \eta\,[\max(0,\,\mathbf{n}_{t,i}\cdot\mathbf{v})]^m$, where $\mathbf{v}\in\mathbb{S}^2$ is the candidate direction and $\eta>0$ is a free scale parameter that absorbs the unknown normalization constant $1/Q_{\max}^m$. The NLS direction estimator is then:
\begin{equation}
(\hat{\mathbf{n}}_{d,\mathrm{NLS}},\,\hat\eta) = \arg\min_{\substack{\mathbf{v}\in\mathbb{S}^2,\;\eta>0 \\ \mathbf{n}_{t,i}\cdot\mathbf{v}\ge0,\;i=1,\ldots,K}} F(\mathbf{v},\eta),
\label{eq:argmin_NL}
\end{equation}
with cost function:
\begin{equation}
F(\mathbf{v},\eta) = \sum_{i=1}^{K}\bigl(\eta\,[\mathbf{n}_{t,i}\cdot\mathbf{v}]^m - p_i\bigr)^2.
\label{eq:NLS_cost}
\end{equation}
To solve~\eqref{eq:argmin_NL}, we reparameterize the direction in spherical coordinates $(\theta_d,\phi_d)$ so that $\mathbf{v}=[\sin\theta_d\cos\phi_d,\;\sin\theta_d\sin\phi_d,\;-\cos\theta_d]^{\mathsf T}$, which satisfies $\lVert\mathbf{v}\rVert=1$ by construction and reduces the parameter vector to $(\theta_d,\phi_d,\eta)\in\mathbb{R}^3$. The sum-of-squares structure of~\eqref{eq:NLS_cost} is then exploited by a Levenberg--Marquardt solver (MATLAB's \texttt{lsqnonlin}), which uses the Gauss--Newton Hessian approximation $\mathbf{J}^{\mathsf T}\mathbf{J}$ and achieves quadratic convergence near the zero-residual solution.

\subsection{Consistency and Optimality at the True Direction}
At the true direction $\mathbf{v}=\mathbf{n}_d$, we have $\mathbf{n}_{t,i}\cdot\mathbf{v}=Q_i$ and hence $[\mathbf{n}_{t,i}\cdot\mathbf{v}]^m=Q_i^m$. From~\eqref{eq:p_target_noiseless}, $p_i=Q_i^m/Q_{\max}^m$. Substituting into~\eqref{eq:NLS_cost}:
\begin{equation}
F(\mathbf{n}_d,\eta)=\sum_{i=1}^K Q_i^{2m}\bigl(\eta - 1/Q_{\max}^m\bigr)^2 = 0
\end{equation}
when $\eta^*=1/Q_{\max}^m$. Thus, in the absence of noise, the NLS achieves a global minimum of zero at the true direction, confirming consistency.

Since $\mathbf{n}_r$ does not appear in~\eqref{eq:p_target}--\eqref{eq:NLS_cost}, the NLS direction estimate is independent of the receiver orientation by construction. The common factor~$\eta$ is eliminated through max-normalization and the free scale parameter~$\hat\eta$, which jointly marginalize over the unknown multiplicative term.

\begin{figure}[tb]
    \centering
    \includegraphics[width=\linewidth]{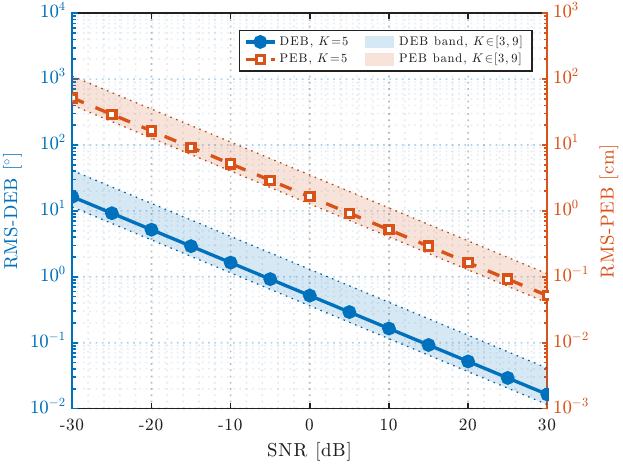}
    \caption{RMS-DEB (solid, left axis) and RMS-PEB (dashed, right axis) versus SNR for the DEB-optimized orientation sets ($\Phi_{1/2}=45^\circ$). Bold curves: $K=5$; shaded bands: range over $K\in\{3,\ldots,9\}$; dotted edges: $K=3$ (upper) and $K=9$ (lower). Both bounds decrease as $1/\sqrt{\mathrm{SNR}}$, with the band width reflecting the diminishing incremental gain from additional orientations at high SNR.}
    \label{fig:PEB_vs_noise}
\end{figure}

\section{Linear Estimators}
\label{sec:Lineal}

The present section develops the direction-finding stage with full mathematical rigor and presents two statistically grounded estimators: (i)~generalized least squares, i.e., GLS, which is minimum-variance under AWGN with known covariance, and (ii)~weighted least squares, i.e., WLS, a less expensive approximation that neglects the weak cross–correlation between equations. For implementation details, an end-to-end pseudocode of the full procedure (GLS/WLS direction finding, distance recovery, and 3D position estimation) is provided in Algorithm~\ref{alg:position}.

\subsection{Direction Estimation via GLS}
\label{subsec:gls_direction}

This subsection consolidates the analytical procedure that converts the raw optical powers into a closed-form, statistically efficient estimate of the direction vector $\mathbf{n}_d$. The exposition proceeds from the deterministic constraints, through a rigorous stochastic characterisation of the residuals, to the final GLS solution. 

\subsubsection{Lambertian Power Ratios and Homogeneous Constraints}
Let $\mu_i=\mathbb E\!\bigl[P_{r,i}\bigr]$ be the noise-free average power at orientation~$i$. We define power ratio $\beta_i$ as:
\begin{equation}
  \beta_i=\left(\frac{\mu_i}{\mu_1}\right)^{1/m}, \quad i=2,\ldots,K,
  \label{eq:beta_def_gls}
\end{equation}
and, from~\eqref{eq:Pr_i}, \eqref{eq:hLOS}, \eqref{eq:cosines} and \eqref{eq:constant_C}:
\begin{equation}
\mu_i=-C\,\frac{\bigl(\mathbf{n}_{t,i}\cdot\mathbf{d}\bigr)^{m}
\bigl(\mathbf{n}_r\cdot\mathbf{d}\bigr)}{d^{m+3}}.
\label{eq:mu_i_C}
\end{equation}
The distance $d$ and the receiver-dependent factor $\cos\psi$ appear identically in both $\mu_i$ and $\mu_1$. Their ratio therefore cancels these common factors, yielding $\beta_i = \cos\phi_i / \cos\phi_1$ from \eqref{eq:cosines}, which depends only on the LED orientations and the displacement direction $\mathbf{d}$.

Using \eqref{eq:mu_i_C} in \eqref{eq:beta_def_gls}, each $\beta_i$ imposes the orthogonality condition:
\begin{equation}
  \mathbf a_i\cdot
  \mathbf d=0,
  \quad
  i=2,\ldots,K,
  \label{eq:homog_gls}
\end{equation}
i.e., $\mathbf a_i \perp \mathbf d$, with:
\begin{equation}
\mathbf a_i\!=\!\mathbf n_{t,i}-\beta_i\mathbf n_{t,1}, 
\label{eq:definition_a_i}
\end{equation}
which therefore supplies $K-1$ homogeneous hyperplanes that intersect at the true~$\mathbf d$.
Geometrically, each constraint $\mathbf{a}_i\cdot\mathbf{d}=0$ defines a hyperplane through the origin containing the true direction $\mathbf{d}$. With $K-1$ such hyperplanes (from $K$ orientations), their intersection yields the direction estimate.

\subsubsection{Stochastic Perturbation of the Ratios}
Recall from~\eqref{eq:sampleMean} that the sample mean is:
\begin{equation}
\bar P_{r,i}=\frac1N\sum_{k=1}^NP_{r,i,k}
 \sim\mathcal N(\mu_i,\sigma^{2}/N).
\end{equation}
For notational convenience, define $\hat{\mu}_i \triangleq \bar P_{r,i}$, which is the MVU estimator of $\mu_i$.
Injecting $\hat\mu_i$ into~\eqref{eq:beta_def_gls} yields:
\begin{equation}
  \hat\beta_i
  =g(\hat\mu_1,\hat\mu_i),
  \qquad
  g(x_1,x_2)=\left(\frac{x_2}{x_1}\right)^{\frac{1}{m}}.
\end{equation}
As demonstrated in Appendix \ref{app:beta_lin_gls}, a first-order Taylor expansion of $g$ centred at
$\boldsymbol\mu_i=(\mu_1,\mu_i)^{\mathsf T}$ gives:
\begin{equation}
  \hat\beta_i
  \approx
  \beta_i
  +\frac{\beta_i}{m}
   \!\left(\frac{n_i}{\mu_i}-\frac{n_1}{\mu_1}\right).
  \label{eq:beta_lin_gls}
\end{equation}
Hence every $\hat\beta_i$ shares the common perturbation~$n_1$,
producing correlation across orientations.

Let us now define the ratio error:
\begin{equation}
\tilde{n}_i=\hat\beta_i-\beta_i.
\label{eq:ratio_error}
\end{equation}
Then, as demonstrated in Appendix \ref{app:cov_beta_gls}, using~\eqref{eq:beta_lin_gls}
and the independence of $n_i$ and $n_1$ leads to:
\begin{align}
  \operatorname{Var}[\tilde{n}_i]
  &=
  \frac{\sigma^{2}}{Nm^{2}}\,
  \beta_i^{2}\bigl(\mu_i^{-2}+\mu_1^{-2}\bigr),
  \label{eq:var_beta_gls}\\
  \operatorname{Cov}[\tilde{n}_i,\tilde{n}_j]
  &=
  \frac{\sigma^{2}}{Nm^{2}}\,
  \beta_i\beta_j\,\mu_1^{-2},
  \quad i\neq j.
  \label{eq:cov_beta_gls}
\end{align}
The factor $(\mu_i^{-2} + \mu_1^{-2})$ in~\eqref{eq:var_beta_gls}
demonstrates that the variance of $\tilde{n}_i$ depends explicitly on
$\mu_i$, even though each $n_i$ has the same variance
$\sigma^2$.  In other words, a single‐LED system exhibits
heteroscedastic ratio errors.

\subsubsection{Residuals and Their Covariance}
Define the empirical constraint normals as
\(\hat{\mathbf a}_i = \mathbf n_{t,i} - \hat\beta_i\,\mathbf n_{t,1}\)
and the corresponding residuals as:
\begin{equation}
\xi_i = \hat{\mathbf a}_i\cdot\mathbf d,\qquad i=2,\ldots,K.
\label{eq:residual_def}
\end{equation}
Using \eqref{eq:ratio_error} and \eqref{eq:definition_a_i}, we have
\(
\hat{\mathbf a}_i=\mathbf a_i-\tilde n_i\,\mathbf n_{t,1}
\).
Since \(\mathbf a_i\cdot\mathbf d=0\) by \eqref{eq:homog_gls}, it follows that:
\begin{equation}
\xi_i
= -\bigl(\mathbf n_{t,1}\!\cdot\!\mathbf d\bigr)\,\tilde n_i
= -\,\kappa\,\tilde n_i,
\qquad
\kappa = \mathbf n_{t,1}\!\cdot\!\mathbf d.
\label{eq:residual_scalar}
\end{equation}
Stacking \(\boldsymbol{\xi}=[\xi_2,\ldots,\xi_K]^{\mathsf T}\) and
\(\tilde{\boldsymbol n}=[\tilde n_2,\ldots,\tilde n_K]^{\mathsf T}\) yields:
\begin{equation}
\begin{aligned}
\boldsymbol{\xi} &= -\,\kappa\,\tilde{\boldsymbol n}, \\
\mathbb E[\boldsymbol{\xi}] &= \mathbf{0}_{(K-1)\times 1}, \\
\mathbf \Sigma_{\boldsymbol{\xi}} &= \operatorname{Cov}(\boldsymbol{\xi})
= \kappa^{2}\mathbf\Sigma_{\beta},
\end{aligned}
\label{eq:residual_cov}
\end{equation}
where \(\mathbf\Sigma_{\beta}\in\mathbb R^{(K-1)\times(K-1)}\) collects the variances and covariances of \(\tilde n_i\) given in \eqref{eq:var_beta_gls}–\eqref{eq:cov_beta_gls}. 

\subsubsection{GLS Formulation and Closed-Form Solution}
Stack the empirical constraint normals into
\(\hat{\mathbf A}=[\hat{\mathbf a}_2\,\cdots\,\hat{\mathbf a}_K]\in\mathbb R^{3\times(K-1)}.\)
From \eqref{eq:residual_def}–\eqref{eq:residual_cov}, the stacked residuals satisfy
\(\boldsymbol{\xi}=\hat{\mathbf A}^{\mathsf T}\mathbf d\)
with
\(\boldsymbol{\xi}\sim\mathcal N(\mathbf{0}_{(K-1)\times 1},\mathbf\Sigma_{\boldsymbol{\xi}})\)
and
\(\mathbf\Sigma_{\boldsymbol{\xi}}=\kappa^{2}\mathbf\Sigma_{\beta}\),
\(\kappa=\mathbf n_{t,1}\!\cdot\!\mathbf d\).
Given an observation of \(\boldsymbol{\xi}=\hat{\mathbf A}^{\mathsf T}\mathbf d\), the maximum-likelihood (equivalently, GLS) estimate of the direction is obtained by minimizing the Mahalanobis norm of \(\boldsymbol{\xi}\) \cite{Bishop2006}:
\begin{equation}
\begin{aligned}
  \hat{\mathbf d}
  &= \arg\min_{\|\mathbf d\|=1}
     \bigl\|\hat{\mathbf A}^{\mathsf T}\mathbf d\bigr\|^{2}_{\mathbf\Sigma_{\boldsymbol{\xi}}^{-1}} \\
  &= \arg\min_{\|\mathbf d\|=1}
     \mathbf d^{\mathsf T}\!\bigl(\hat{\mathbf A}\,\mathbf\Sigma_{\boldsymbol{\xi}}^{-1}\hat{\mathbf A}^{\mathsf T}\bigr)\mathbf{d}.     
\end{aligned}
\label{eq:gls_criterion_final}
\end{equation}
Since \(\mathbf\Sigma_{\boldsymbol{\xi}}=\kappa^{2}\mathbf\Sigma_{\beta}\), the unknown scale \(\kappa^{2}\) cancels in \eqref{eq:gls_criterion_final}; hence one may use \(\mathbf\Sigma_{\beta}^{-1}\) in place of \(\mathbf\Sigma_{\boldsymbol{\xi}}^{-1}\) without affecting the minimizer. The problem is a constrained Rayleigh quotient whose solution is the unit eigenvector associated with the smallest eigenvalue of:
\begin{equation}
  \mathbf M_{\rm GLS}=\hat{\mathbf A}\,\mathbf\Sigma_{\beta}^{-1}\hat{\mathbf A}^{\mathsf T}\in\mathbb R^{3\times3}.
  \label{eq:M_GLS_final}
\end{equation}
Therefore:
\begin{equation}
  \hat{\mathbf n}_{d,{\rm GLS}}
  = \frac{\mathbf v_{\min}\!\bigl(\mathbf M_{\rm GLS}\bigr)}
         {\bigl\|\mathbf v_{\min}\!\bigl(\mathbf M_{\rm GLS}\bigr)\bigr\|},
  \qquad
  \hat{\mathbf n}_{d,{\rm GLS}}\cdot\mathbf n_{t,1} > 0,
  \label{eq:gls_dir_final}
\end{equation}
where \(\mathbf v_{\min}(\cdot)\) denotes the eigenvector corresponding to the smallest eigenvalue; the sign convention ensures the direction points from the transmitter toward the receiver.

\begin{algorithm}[tb]
  \caption{3D Position Estimation from $K$ Orientations}
  \label{alg:position}
  \small
  \begin{algorithmic}[1]
  \Require
    \begin{tabular}[t]{@{}l@{}}
      Power samples $P_{r,i,k}$ ($k=1,\dots,N$, $i=1,\dots,K$);\\
      steering vectors $\mathbf n_{t,i}$; Lambertian order $m$;\\
      noise variance $\sigma^{2}$; \textsc{mode}$\in\{\mathrm{GLS},\mathrm{WLS}\}$;\\
      radiometric constant $C$.
    \end{tabular}
  \Ensure
    $\hat{\mathbf r}$ (3D position of the receiver)
  \Statex
  \State {\bfseries Direction finding:}
  \State \quad Compute $\hat\mu_i=\tfrac{1}{N}\sum_{k=1}^N P_{r,i,k}$
            for $i=1,\dots,K$
  \State \quad Compute $\hat\beta_i=(\hat\mu_i/\hat\mu_1)^{1/m}$ for $i=2,\dots,K$
  \State \quad Form $\mathbf a_i = \mathbf n_{t,i} - \hat\beta_i\,\mathbf n_{t,1}$
  \State \quad Set $\mathbf A = [\mathbf a_2,\dots,\mathbf a_K]$
  \If{\textsc{mode} = GLS}
    \State \quad Build $\mathbf\Sigma_\beta$ via
           \eqref{eq:var_beta_gls}--\eqref{eq:cov_beta_gls},
           substituting $\beta_i \leftarrow \hat\beta_i$
    \State \quad $\mathbf M \gets \mathbf A\,\mathbf\Sigma_\beta^{-1}\,\mathbf A^{\mathsf T}$
  \Else\Comment{WLS (diagonal approximation)}
    \State \quad $\mathbf{\tilde\Sigma}_\beta \gets 
           \operatorname{diag}\bigl\{\hat\beta_i^{2}\bigl(\hat\mu_i^{-2}+\hat\mu_1^{-2}\bigr)\bigr\}$
    \State \quad $\mathbf M \gets \mathbf A\,\mathbf{\tilde\Sigma}_\beta^{-1}\,\mathbf A^{\mathsf T}$
  \EndIf
  \State \quad Compute $\mathbf v_{\min} =$ eigenvector of $M$ with smallest eigenvalue
  \State \quad $\hat{\mathbf n}_d \gets \mathbf v_{\min}/\|\mathbf v_{\min}\|$
  \If{$\hat{\mathbf n}_d\cdot\mathbf n_{t,1} < 0$}
    \State \quad $\hat{\mathbf n}_d \gets -\,\hat{\mathbf n}_d$ \Comment{Ensure pointing to Rx}
  \EndIf

  \Statex
  \State {\bfseries Distance estimation (cooperative alignment):}
  \State \quad Steer LED to $\mathbf n_t \gets \hat{\mathbf n}_d$,\quad reorient PD to $\mathbf n_r \gets -\,\hat{\mathbf n}_d$
  \State \quad Collect $N$ power samples $P_{r,K+1}$ 
  \State \quad Compute $\hat\mu_{K+1} = \tfrac{1}{N}\sum_{k=1}^{N} P_{r,K+1,k}$
  \State \quad $\hat d \gets \sqrt{\tfrac{C}{\hat\mu_{K+1}}}$

  \Statex
  \State {\bfseries Final 3D position:}
  \State \quad $\hat{\mathbf r} \gets \mathbf t + \hat d \,\hat{\mathbf n}_d$
  \State \Return $\hat{\mathbf r}$
  \end{algorithmic}
\end{algorithm}

\subsection{WLS as a Practical Simplification}

The GLS solution in \eqref{eq:gls_criterion_final} requires inverting the \((K\!-\!1)\times(K\!-\!1)\) covariance \(\mathbf\Sigma_{\boldsymbol{\xi}}=\kappa^{2}\mathbf\Sigma_{\beta}\), typically via a Cholesky factorization \cite{golub2013matrix}. For real-time, low-power receivers, a lighter alternative is to neglect the off-diagonal terms in \eqref{eq:cov_beta_gls} and approximate:
\begin{equation}
\tilde{\mathbf\Sigma}_{\beta}
=
\operatorname{diag}\!\Bigl\{
\beta_i^{2}\bigl(\mu_i^{-2}+\mu_1^{-2}\bigr)
\Bigr\}_{i=2}^{K}.
\label{eq:Sigma_beta_diag}
\end{equation}
Since \(\mathbf\Sigma_{\boldsymbol{\xi}}=\kappa^{2}\mathbf\Sigma_{\beta}\), the unknown scale \(\kappa^{2}\) cancels, and we define the WLS weighting matrix as \(\tilde{\mathbf\Sigma}_{\beta}^{-1}\).

Using the empirical normals \(\hat{\mathbf A}\) from the previous subsection, the WLS criterion is the Rayleigh quotient:
\begin{equation}
J_{\rm WLS}(\mathbf d)
=\mathbf d^{\mathsf T}\!\bigl(\hat{\mathbf A}\,\tilde{\mathbf\Sigma}_{\beta}^{-1}\,\hat{\mathbf A}^{\mathsf T}\bigr)\mathbf d,
\qquad \|\mathbf d\|=1,
\label{eq:WLS_cost}
\end{equation}
and we introduce the associated WLS matrix:
\begin{equation}
\mathbf M_{\rm WLS}
=
\hat{\mathbf A}\,\tilde{\mathbf\Sigma}_{\beta}^{-1}\,\hat{\mathbf A}^{\mathsf T}\in\mathbb R^{3\times 3}.
\label{eq:M_WLS}
\end{equation}
The minimizer is the unit eigenvector corresponding to the smallest eigenvalue of \(\mathbf M_{\rm WLS}\):
\begin{equation}
\hat{\mathbf n}_{d,{\rm WLS}}
=\frac{\mathbf v_{\min}\!\bigl(\mathbf M_{\rm WLS}\bigr)}
      {\bigl\|\mathbf v_{\min}\!\bigl(\mathbf M_{\rm WLS}\bigr)\bigr\|},
\qquad
\hat{\mathbf n}_{d,{\rm WLS}}\cdot\mathbf n_{t,1} > 0.
\label{eq:WLS_solution}
\end{equation}

The neglected cross-covariances in \eqref{eq:cov_beta_gls} scale with \(\mu_1^{-2}\), whereas the retained diagonal terms scale with \(\mu_i^{-2}+\mu_1^{-2}\). Thus, when the reference orientation \(i=1\) is chosen such that \(\mu_1\gg \mu_i\) for most \(i\), the off-diagonal terms are negligible relative to the diagonal terms and WLS closely tracks GLS while avoiding a full covariance inversion. In more symmetric geometries where \(\mu_i\approx\mu_1\) for many \(i\), the cross-correlations are non-negligible and the full GLS should be used for optimal performance.

\subsection{Receiver-Orientation Independence}
\label{subsec:nr_independence}
Having defined all three direction estimators---NLS in Section~\ref{sec:NoLineal}, and GLS/WLS above---we now establish a property common to all of them: the direction estimates are independent of the receiver orientation $\mathbf{n}_r$. From~\eqref{eq:mu_factorized}--\eqref{eq:eta_nuisance}, the channel model factorizes as $\mu_i = \eta\,Q_i^m$, where $\eta = C\cos\psi/d^2$ is a positive scalar common to all $K$ orientations, since the receiver remains stationary throughout the $K$-orientation acquisition.

For GLS and WLS, the power ratios $\beta_i = (\mu_i/\mu_1)^{1/m} = Q_i/Q_1$~\eqref{eq:beta_def_gls} cancel $\eta$ exactly. Since the constraint vectors $\mathbf{a}_i$~\eqref{eq:definition_a_i}, the covariance $\mathbf{\Sigma}_\beta$~\eqref{eq:var_beta_gls}--\eqref{eq:cov_beta_gls}, and the matrices $\mathbf{M}_{\mathrm{GLS}}$~\eqref{eq:M_GLS_final} and $\mathbf{M}_{\mathrm{WLS}}$~\eqref{eq:M_WLS} are all constructed from $\{\beta_i\}$ and $\{\mathbf{n}_{t,i}\}$, the direction estimates are $\mathbf{n}_r$-independent. The GLS weighting via $\mathbf{\Sigma}_\beta^{-1}$ further ensures that hyperplanes~\eqref{eq:homog_gls} with lower noise variance contribute more to the eigenvector solution.

For NLS, the normalized targets $p_i = \mu_i/\max_j\mu_j = Q_i^m/Q_{\max}^m$~\eqref{eq:p_target_noiseless} cancel $\eta$ through a different mechanism: the free scale parameter $\hat\eta$ absorbs the unknown common factor during optimization (cf.\ Section~\ref{sec:NoLineal}).

This property holds regardless of whether $\mathbf{n}_r$ is known or unknown, and numerical confirmation under random receiver tilts is reported in Section~\ref{sec:results}.

\section{Estimators Comparison}
\label{sec:results}

\subsection{Direction-Finding Performance}
\label{subsec:DF_performance}

This subsection evaluates the direction-finding stage in isolation, using the angular error $\theta_{\mathrm{err}} = \arccos(\hat{\mathbf{n}}_d\cdot\mathbf{n}_d)$ as the performance metric, independently of distance recovery. For each transmitter orientation $i$, the received optical power was simulated according to \eqref{eq:Pr_i} and \eqref{eq:hLOS} using the system parameters in Table~\ref{tab:ga_params}, with a semi-angle at half power $\Phi_{1/2}=45^\circ$. Zero-mean AWGN $n_i\!\sim\!\mathcal{N}(0,\sigma^2)$ was drawn i.i.d.\ across orientations and samples; the sample mean $\bar P_{r,i}$ was formed from $N$ realizations as in \eqref{eq:sampleMean}. For each estimator, $M=1{,}000$ Monte Carlo trials were performed at each of the $|\mathcal{R}|=1{,}792$ testbed positions for $K=5$ orientations. To ensure a fair comparison, each estimator uses its best GA-optimized orientation set: GLS and WLS use the set optimized with the GLS angular RMSE as cost function, while NLS uses the set optimized with the DEB as cost function.

\subsubsection{Angular Error Cumulative Distribution Function}

Table~\ref{tab:DF_performance} and Fig.~\ref{fig:NR_robust_CDF} (solid curves) report the spatial RMSE, the 90th percentile of the error cumulative distribution function (CDF), denoted as $\mathrm{CDF}_{90\%}$, and the mean angular error. The hierarchy DEB $<$ NLS $<$ GLS $<$ WLS is consistent with theory: the DEB ($0.52^\circ$) is the Cramér--Rao bound; NLS ($0.63^\circ$) operates on the full nonlinear Lambertian model without linearization; GLS ($0.66^\circ$) linearizes the ratio model but uses the optimal $\mathbf{\Sigma}_\beta^{-1}$ weighting; and WLS ($0.77^\circ$) further simplifies to diagonal weights. All estimators achieve sub-degree RMSE for direction finding.

\begin{table}[t]
\centering
\caption{Direction-finding angular error ($K=5$, $M=1{,}000$ trials).}
\label{tab:DF_performance}
\begin{tabular}{|l|c|c|c|}
\hline
\textbf{Method} & \textbf{RMSE [$^\circ$]} & $\textbf{CDF}_{90\%}$ \textbf{[$^\circ$]} & \textbf{Mean [$^\circ$]} \\ \hline
DEB   & 0.52 & 0.75 & 0.46 \\ \hline
NLS   & 0.63 & 0.81 & 0.52 \\ \hline
GLS   & 0.66 & 0.99 & 0.54 \\ \hline
WLS   & 0.77 & 1.14 & 0.61 \\ \hline
\end{tabular}
\end{table}

\subsubsection{Robustness to Receiver Tilt}

To numerically validate the $\mathbf{n}_r$-independence established in Section~\ref{subsec:nr_independence}, we evaluate direction-finding performance under random receiver tilts. Each tilted receiver normal is parameterized as $\mathbf{n}_r = [\sin\theta_{\mathrm{tilt}}\cos\varphi_{\mathrm{tilt}},\;\sin\theta_{\mathrm{tilt}}\sin\varphi_{\mathrm{tilt}},\;\cos\theta_{\mathrm{tilt}}]^{\mathsf T}$, where $\theta_{\mathrm{tilt}}$ is the polar angle between $\mathbf{n}_r$ and the vertical axis $\mathbf{u}_z$ and $\varphi_{\mathrm{tilt}}$ is the azimuth in the horizontal plane. The tilt angle $\theta_{\mathrm{tilt}}$ is drawn from a half-normal distribution with $\sigma=5^\circ$ truncated at $30^\circ$, and $\varphi_{\mathrm{tilt}}$ is uniform on $[0^\circ,360^\circ)$. For each of the $1{,}792$ positions, $N_{\mathrm{tilt}}=100$ independent random tilts are generated, with $M=1{,}000$ noise trials per tilt realization.

\begin{figure}[tb]
    \centering
    \includegraphics[width=\linewidth]{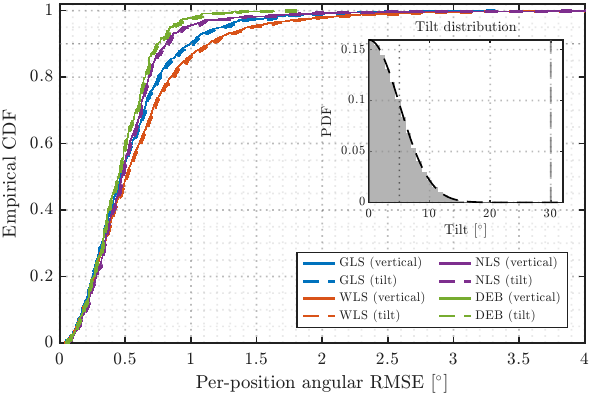}
    \caption{Empirical CDF of the per-position angular RMSE for $K=5$ orientations ($M=1{,}000$ noise trials, $1{,}792$ positions). Solid curves: vertical receiver ($\mathbf{n}_r=[0,0,1]^{\mathsf T}$); dashed curves: random receiver tilt ($N_{\mathrm{tilt}}=100$ realizations per position). Each estimator uses its GA-optimized orientation set. Inset: truncated half-normal tilt distribution ($\sigma=5^{\circ}$, $\theta_{\max}=30^{\circ}$). The near-perfect overlap between solid and dashed curves confirms that receiver-orientation variations have negligible impact on direction-finding accuracy (Section~\ref{subsec:nr_independence}).}
    \label{fig:NR_robust_CDF}
\end{figure}

Table~\ref{tab:tilt_robustness} reports the RMSE degradation relative to the vertical baseline ($\mathbf{n}_r=[0,0,1]^{\mathsf T}$). All estimators exhibit degradation below~$3\%$, confirming that receiver-orientation variations have negligible impact on direction-finding accuracy. This residual degradation is not intrinsic to the estimators but is attributable to SNR reduction---a tilted PD intercepts less optical power ($\cos\psi<1$), lowering the effective SNR---rather than structural bias. Figure~\ref{fig:NR_robust_CDF} corroborates these results: the empirical CDFs of the per-position angular RMSE under random tilt (dashed curves) are virtually indistinguishable from those obtained with a vertical receiver (solid curves) for all four methods, with the inset depicting the truncated half-normal tilt distribution applied in the simulation.

\begin{table}[t]
\centering
\caption{Direction-finding RMSE under random receiver tilt ($K=5$, half-normal $\sigma=5^\circ$, $\max=30^\circ$).}
\label{tab:tilt_robustness}
\begin{tabular}{|l|c|c|c|}
\hline
\textbf{Method} & \textbf{Baseline [$^\circ$]} & \textbf{Random tilt [$^\circ$]} & \textbf{Degradation} \\ \hline
GLS   & 0.657 & 0.671 & $+2.24\%$ \\ \hline
WLS   & 0.766 & 0.786 & $+2.62\%$ \\ \hline
NLS   & 0.620 & 0.631 & $+1.68\%$ \\ \hline
DEB   & 0.518 & 0.527 & $+1.69\%$ \\ \hline
\end{tabular}
\end{table}

\subsubsection{Computational Complexity and Latency}
\label{subsec:complexity}

We summarize the per-inference computational cost for direction finding under $K$ LED orientations and $N$ samples per orientation. Computing the sample means $\{\hat{\mu}_i\}_{i=1}^{K}$ via \eqref{eq:sampleMean} incurs $\mathcal{O}(NK)$ and is common to all schemes.

\paragraph*{GLS}
From \eqref{eq:var_beta_gls}--\eqref{eq:cov_beta_gls}, the full covariance
$\mathbf{\Sigma}_\beta\in\mathbb{R}^{(K-1)\times(K-1)}$ is assembled in $\mathcal{O}(K^2)$, then inverted and used to form $\mathbf{M}_{\rm GLS}=\hat{\mathbf A}\,\mathbf{\Sigma}_\beta^{-1}\hat{\mathbf A}^{\mathsf T}$ in \eqref{eq:M_GLS_final}. The inversion dominates with $\mathcal{O}((K{-}1)^3)$ flops; the subsequent matrix products are $\mathcal{O}(K^2)$, and the $3\times3$ eigenproblem is constant-cost. Overall: $\mathcal{O}(NK)+\mathcal{O}(K^3)$.

\paragraph*{WLS}
Replacing $\mathbf{\Sigma}_\beta$ by the diagonal $\tilde{\mathbf{\Sigma}}_\beta$ in \eqref{eq:Sigma_beta_diag} removes the full covariance inversion. Forming $\mathbf{M}_{\rm WLS}=\hat{\mathbf A}\,\tilde{\mathbf{\Sigma}}_\beta^{-1}\hat{\mathbf A}^{\mathsf T}$ requires only column scalings and $K{-}1$ rank-one updates of a $3\times3$ matrix, i.e., $\mathcal{O}(K)$. The eigenproblem is again constant-cost. Overall: $\mathcal{O}(NK)+\mathcal{O}(K)$.

\paragraph*{NLS}
Solving \eqref{eq:argmin_NL} via the spherical reparameterization described above yields a three-variable unconstrained least-squares problem solved by MATLAB's \texttt{lsqnonlin} (Levenberg--Marquardt). At each iteration, the $K$-dimensional residual vector and its $K{\times}3$ Jacobian are evaluated in $\mathcal{O}(K)$, and the $3{\times}3$ normal equation is solved in constant time. With $T_{\rm NLS}$ Levenberg--Marquardt iterations, the per-estimate cost is $\mathcal{O}(NK)+\mathcal{O}(T_{\rm NLS}\,K)$.

\begin{figure}[tb]
    \centering
    \includegraphics[width=\linewidth]{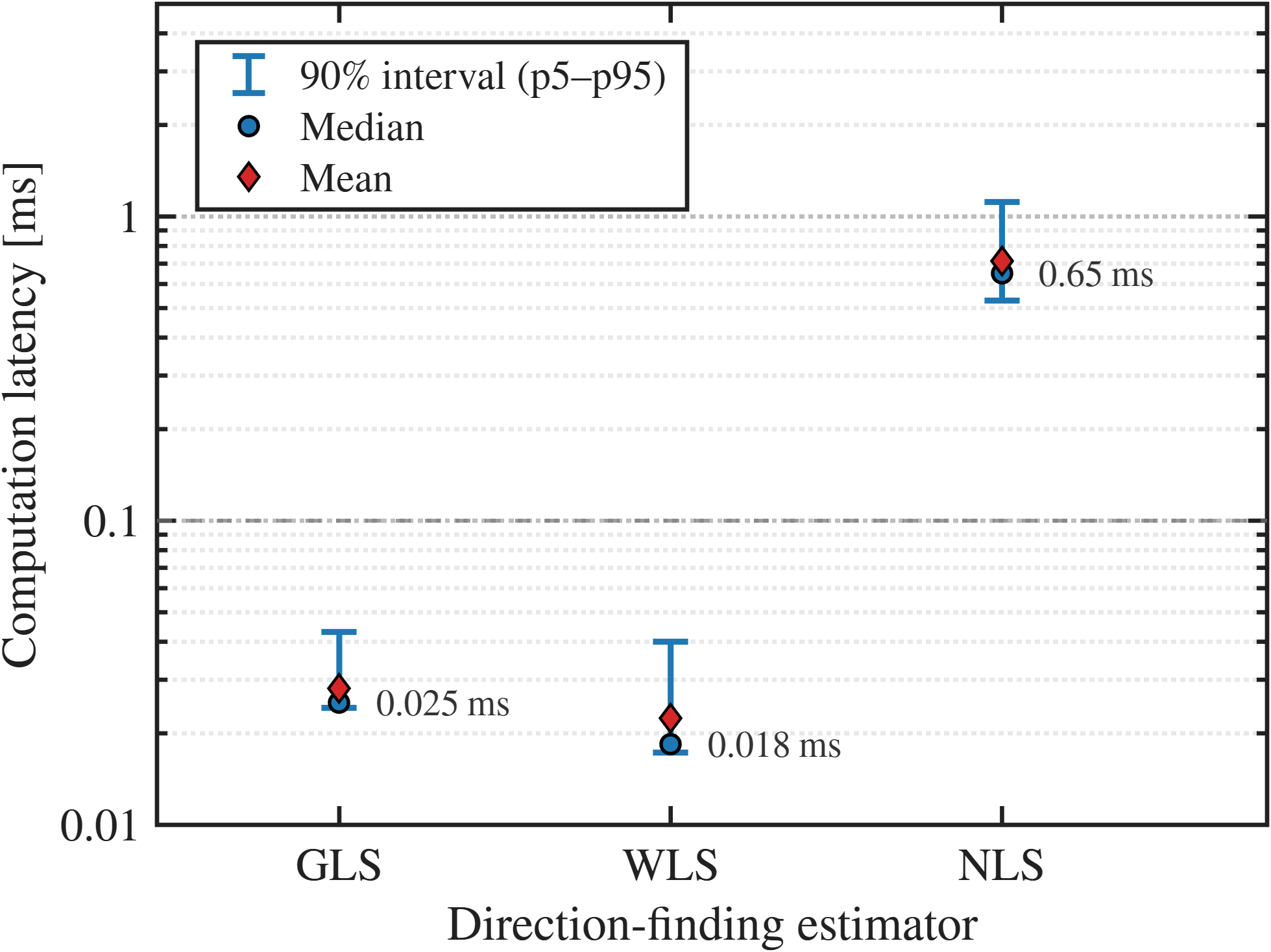}
    \caption{Computation latency of the GLS, WLS, and NLS direction-finding estimators for $K=5$. Markers denote the median and mean over all testbed positions; whiskers span the 90\% interval (p5--p95). The dashed line marks the $0.1$\,ms threshold.}
    \label{fig:latency}
\end{figure}

Figure~\ref{fig:latency} reports the measured computation latency per direction estimate for $K=5$ (powers pre-acquired). WLS and GLS complete with median latencies of $0.018$\,ms and $0.025$\,ms, respectively, consistent with their closed-form $\mathcal{O}(K)$ and compact $\mathcal{O}(K^3)$ structures acting on a small $(K{-}1)\times(K{-}1)$ system followed by a constant-cost $3\times3$ eigensolve. NLS requires a median of $0.65$\,ms---roughly one order of magnitude slower---due to the iterative Levenberg--Marquardt solve. The 90\% intervals (p5--p95) shown in the figure further reveal that NLS exhibits moderate variability ($0.53$--$1.11$\,ms), whereas GLS and WLS remain tightly concentrated below $0.05$\,ms. These measurements corroborate the complexity analysis: closed-form WLS/GLS enable real-time direction finding without fingerprinting or model retraining, whereas NLS incurs higher but still sub-millisecond latency.

\subsection{3D Positioning Performance}
\label{subsec:performance}

\begin{figure}[tb]
    \centering
    \includegraphics[width=\linewidth]{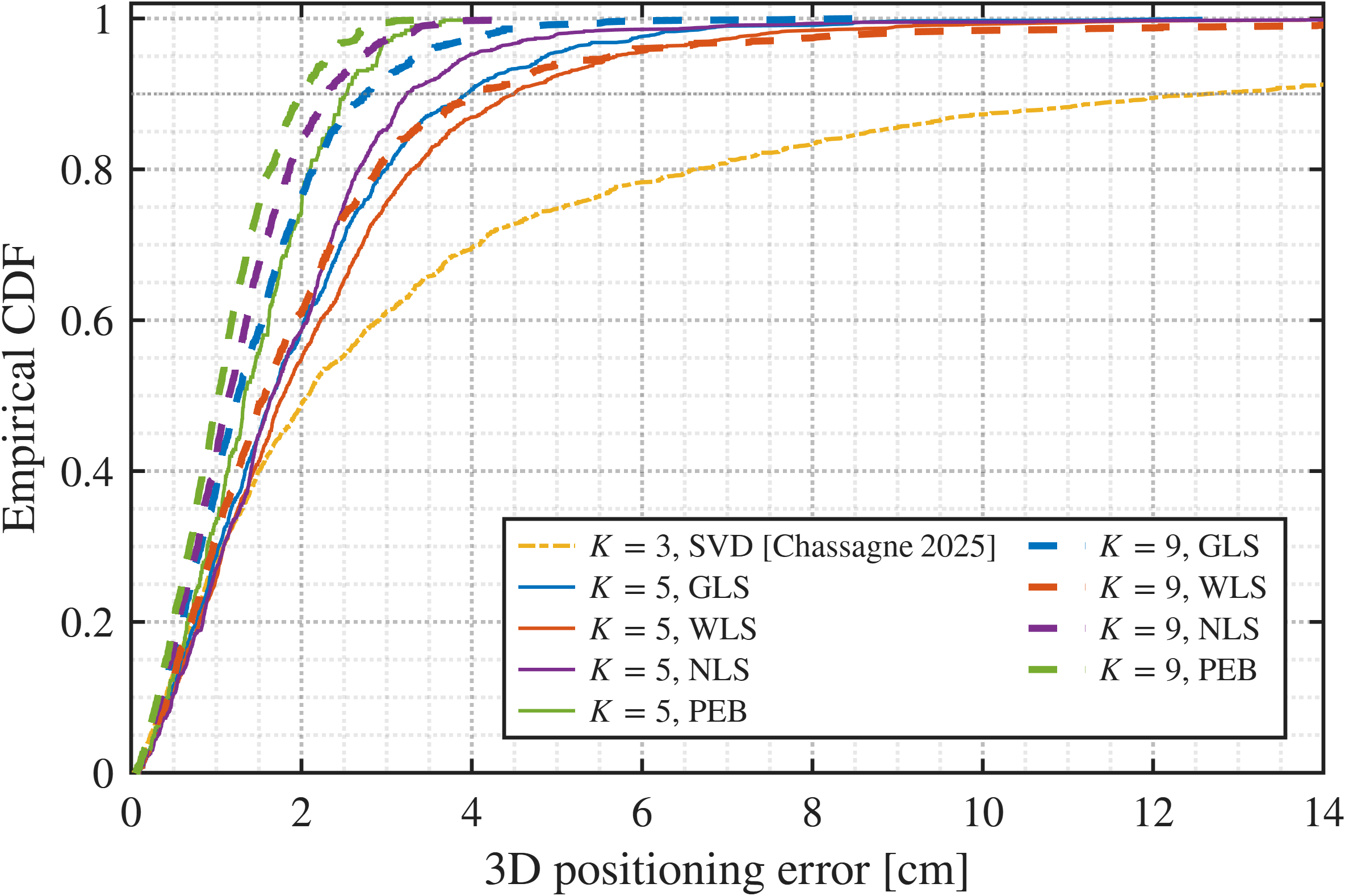}
    \caption{Empirical CDF of the 3D positioning error for $K\in\{3,5,9\}$ ($M=1{,}000$ Monte Carlo trials, $1{,}792$ testbed positions). Solid lines: $K=5$; dashed lines: $K=9$; dash-dotted: $K=3$ SVD baseline~\cite{Chassagne2025}. Each estimator uses its respective GA-optimized orientation set.}
    \label{fig:CDF}
\end{figure}

This subsection extends the evaluation to the full two-stage pipeline (direction finding followed by distance recovery), using the same simulation setup described in Section~\ref{subsec:DF_performance}. Position estimates were computed over the 3D testbed specified in Table~\ref{tab:ga_params}, using the GA-optimized orientation sets detailed below. As in the direction-finding evaluation (Section~\ref{subsec:DF_performance}), each estimator uses its best orientation set: GLS and WLS employ the direction-finding set optimized for GLS, while NLS uses the DEB-optimized set; the PEB is also evaluated with the DEB-optimized orientations. The 3D Euclidean position error was recorded for each of $M=1{,}000$ trials at each of the $|\mathcal{R}|=1{,}792$ testbed positions. Figure~\ref{fig:CDF} plots the empirical CDF of the per-position 3D positioning error for $K\in\{3,5,9\}$. The case $K=3$ employs the deterministic singular value decomposition (SVD) baseline of~\cite{Chassagne2025}, originally proposed for 2D estimation; by applying the SVD to the matrix formed by the vectors in \eqref{eq:definition_a_i}, we extend it to 3D, albeit restricted to three orientations and without a weighted formulation. The setting $K=5$ is selected as a latency--accuracy compromise based on the PEB analysis, and $K=9$ represents the upper bound of the orientation counts studied herein. For GLS, NLS, and the PEB, the $K=9$ configurations (dashed traces) stochastically dominate their $K=5$ counterparts (solid lines), in agreement with the PEB trend that additional orientations reduce the positioning bound. All proposed estimators substantially outperform the $K=3$ baseline~\cite{Chassagne2025}.

Table~\ref{tab:estimation_performance} summarizes the quantitative positioning metrics: RMSE, $\mathrm{CDF}_{90\%}$, and the average positioning error (APE). For $K=5$, NLS achieved the lowest estimator RMSE of $2.40$~cm and $\mathrm{CDF}_{90\%}=3.23$~cm, followed by GLS ($2.52$~cm) and WLS ($2.92$~cm). GLS remained within a factor $\approx1.54\times$ of the PEB ($1.64$~cm RMSE). Increasing to $K=9$ delivered gains for GLS and NLS: GLS improved to $1.78$~cm (a $29\%$ reduction) and NLS to $1.47$~cm (a $39\%$ reduction), narrowing the gap to the PEB ($1.28$~cm). WLS, however, degraded from $2.92$~cm at $K=5$ to $3.42$~cm at $K=9$; this behavior is consistent with its diagonal covariance approximation becoming increasingly suboptimal as the number of orientations grows and the off-diagonal terms neglected in the WLS simplification become non-negligible.
The performance gap between the estimators and the PEB arises from two factors: (i)~the two-stage architecture discards distance information contained in the absolute power levels during direction finding, whereas the PEB assumes joint estimation from all $K+1$ measurements; and (ii)~the PEB assumes perfect beam alignment in the distance-recovery measurement, which no practical estimator can achieve.
The APE values track the same ordering as RMSE and \(\mathrm{CDF}_{90\%}\), indicating that the improvements are uniform across the distribution rather than confined to the median or the tail.

\begin{table}[t]
\centering
\caption{Performance of position-estimation\\ methods in a \(3\times3\times2\,\mathrm{m^3}\) room.}
\label{tab:estimation_performance}
\begin{tabular}{|c| c |c |c| c|}
\hline
\textbf{$K$} & \textbf{Method} & \textbf{RMSE [cm]} & $\textbf{CDF}_{90\%}$ \textbf{[cm]} & \textbf{APE [cm]} \\ \hline
3 & \cite{Chassagne2025} & 11.05 & 12.61 &  5.35 \\ \hline
5 & GLS   &  2.52 &  3.93 &  \textbf{2.00} \\ \hline
5 & WLS   &  2.92 &  4.49 &  2.25 \\ \hline
5 & NLS   &  2.40 &  3.23 &  1.90 \\ \hline
5 & PEB   &  1.64 &  2.53 &  1.43 \\ \hline
9 & GLS   &  1.78 &  2.75 &  \textbf{1.46} \\ \hline
9 & WLS   &  3.42 &  4.01 &  2.15 \\ \hline
9 & NLS   &  1.47 &  2.28 &  1.26 \\ \hline
9 & PEB   &  1.28 &  2.00 &  1.10 \\ \hline
\end{tabular}
\\
\vspace{4pt}
\footnotesize \textit{Note}: For the PEB rows, ``RMSE'' denotes the RMS of the PEB over all testbed positions (a deterministic spatial statistic, not a Monte Carlo estimate); ``CDF$_{90\%}$'' is the 90th percentile of the PEB distribution across positions; ``APE'' is the spatial mean PEB.
\end{table}

\begin{figure}[tb]
    \centering
    \includegraphics[width=\linewidth]{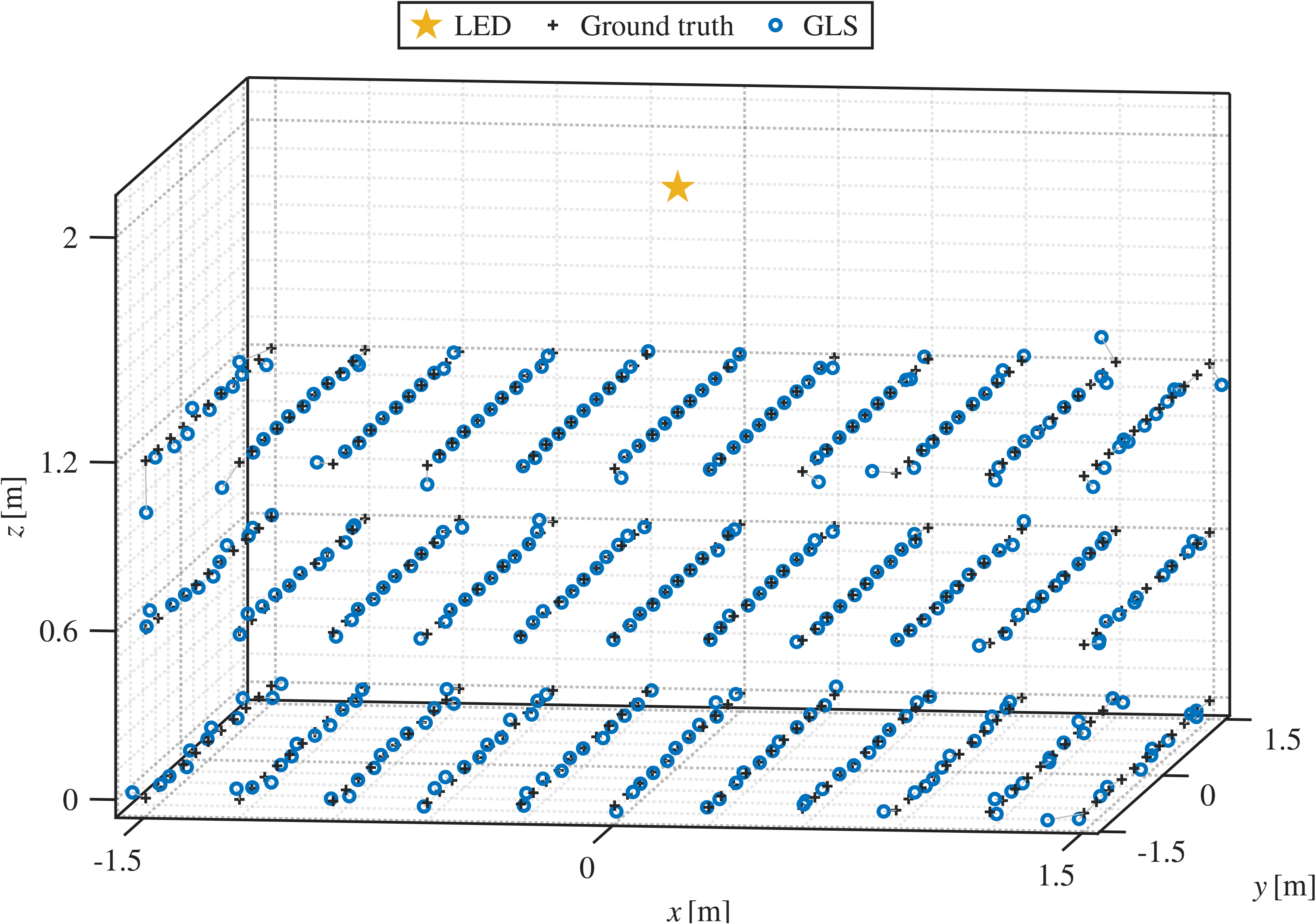}
    \caption{3D position estimates produced by GLS ($K=5$) at receiver heights $z\in\{0,\,0.6,\,1.2\}\,\mathrm{m}$ in the $3\times3\times2\,\mathrm{m^3}$ room. The LED transmitter is at the ceiling ($z=2\,\mathrm{m}$); segments connect each reference position to the corresponding estimate.}
    \label{fig:estimation}
\end{figure}

Figure~\ref{fig:estimation} depicts the 3D position estimates produced by GLS for $K=5$ at three receiver heights $z\in\{0,\,0.6,\,1.2\}\,\mathrm{m}$, with the LED transmitter located at the ceiling ($z=2\,\mathrm{m}$). The displacement between each reference position and the corresponding estimate reveals the spatial structure of the positioning error. At $z=1.2\,\mathrm{m}$, errors grow towards the testbed boundaries---particularly at the corners---consistent with the DEB heat maps in Fig.~\ref{fig:optimo_vs_random_3D}(a) and attributable to the reduced SNR at large incidence angles. At lower heights ($z=0$ and $z=0.6\,\mathrm{m}$), the estimates remain tightly concentrated around the reference positions, in agreement with the low RMSE reported in Table~\ref{tab:estimation_performance}.

\subsection{Estimator Performance vs.\ SNR}
\label{subsec:vs_snr}

\begin{figure}[tb]
    \centering
    \subfloat[]{\includegraphics[width=0.9\linewidth]{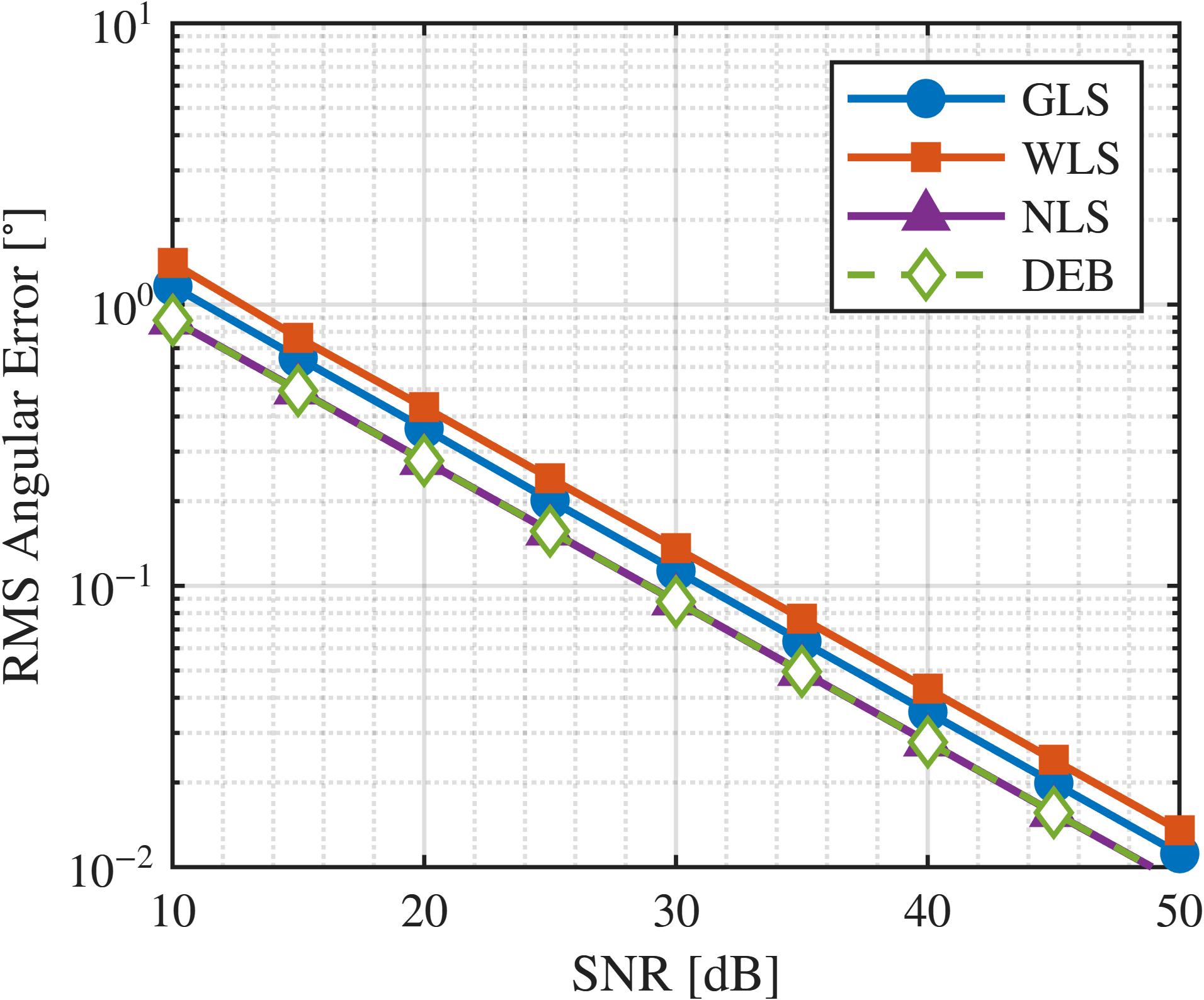}%
    \label{fig:rmse_vs_snr_DF}}\\
    \subfloat[]{\includegraphics[width=0.9\linewidth]{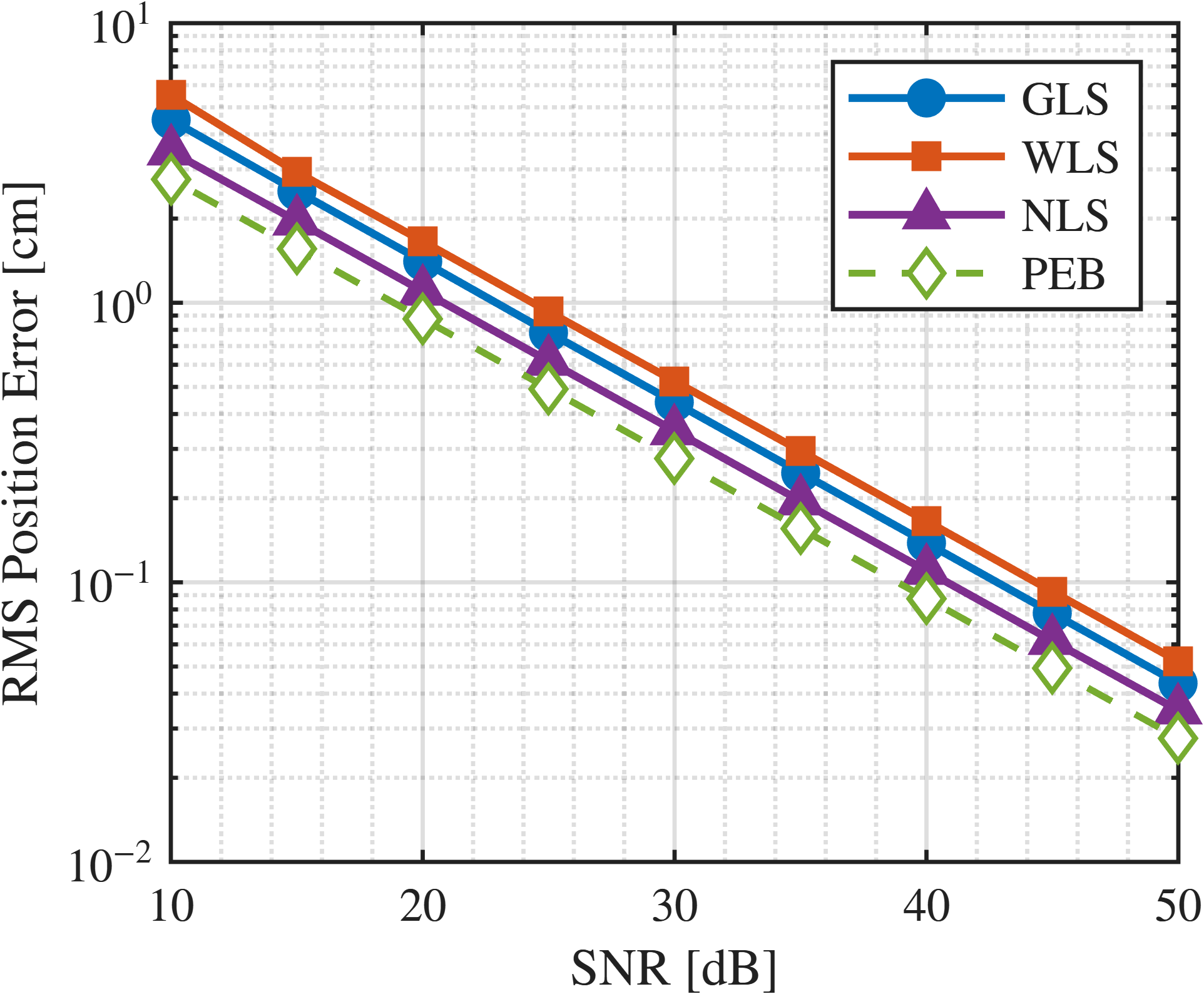}%
    \label{fig:rmse_vs_snr_3D}}
    \caption{Realized estimator performance vs.\ SNR for $K=5$ ($M=1{,}000$ Monte Carlo trials, $1,792$ testbed positions). (a)~RMS angular error (direction finding): NLS coincides with the DEB; GLS and WLS remain at constant factors ${\approx}1.3{\times}$ and ${\approx}1.6{\times}$. (b)~RMS 3D positioning error: all estimators track the PEB with the same constant-factor offsets. All curves follow the $1/\sqrt{\mathrm{SNR}}$ slope.}
    \label{fig:rmse_vs_snr}
\end{figure}

Whereas Table~\ref{tab:estimation_performance} reports performance at the nominal operating point ($\mathrm{SNR}=14$\,dB), this subsection characterizes how the estimators scale with the noise level. Unlike the theoretical bounds in Fig.~\ref{fig:PEB_vs_noise}, the curves in Fig.~\ref{fig:rmse_vs_snr} represent the realized performance of each estimator obtained from full Monte Carlo simulation ($K=5$, $M=1{,}000$ trials per position, $|\mathcal{R}| = 1,792$ testbed positions, $\mathrm{SNR}\in[10,50]$\,dB). This provides the practical counterpart to the bound-only analysis: while Fig.~\ref{fig:PEB_vs_noise} shows the best achievable performance, Fig.~\ref{fig:rmse_vs_snr} shows what the proposed estimators actually attain.

All estimator curves in Fig.~\ref{fig:rmse_vs_snr} decrease with the $1/\sqrt{\mathrm{SNR}}$ slope predicted by Cramér--Rao theory, confirming that the proposed estimators operate in the asymptotic regime across the entire SNR range. For direction finding (Fig.~\ref{fig:rmse_vs_snr}\,(a)), NLS tracks the DEB within a factor ${\approx}1.00$, indicating near-efficient estimation; GLS and WLS remain at constant factors of ${\approx}1.3{\times}$ and ${\approx}1.6{\times}$ above the DEB, respectively, consistent with the suboptimality inherent in the linearized ratio model. For 3D positioning (Fig.~\ref{fig:rmse_vs_snr}\,(b)), the same ordering holds: NLS achieves the lowest RMSE (e.g., $1.10$\,cm at $20$\,dB vs.\ PEB $0.88$\,cm), followed by GLS ($1.41$\,cm) and WLS ($1.66$\,cm). These constant-factor gaps persist across all SNR values, confirming that the performance differences are structural rather than noise-regime-dependent. The SNR axis can also be interpreted in terms of ambient light: higher ambient illuminance increases the shot-noise contribution to $\sigma^2$, reducing the effective SNR and shifting the operating point leftward along the curves.

These results highlight a practical trade-off: NLS achieves the best accuracy (nearly matching the bound) but requires ${\approx}0.65$\,ms per estimate due to its iterative Levenberg--Marquardt solver, whereas GLS and WLS---despite operating $1.3$--$1.6{\times}$ above the bound---complete in ${\approx}0.025$\,ms and $0.018$\,ms, respectively, thanks to their closed-form structure. For latency-critical applications, GLS offers the best accuracy among the closed-form estimators with microsecond-level latency. For scenarios where sub-millisecond latency is acceptable, NLS delivers near-optimal performance.

\section{Conclusion and Future Works}
\label{sec:Conclusion}

This paper introduced a beam–steered single-LED, single-PD OWP architecture that delivers full 3D indoor localization without receiver rotation, cameras, or PD arrays. Our method is explicitly model-based, grounded in radiometry and geometry, so it requires no fingerprinting or data retraining, enabling interpretable performance, principled uncertainty analysis, and efficient co-design of hardware and algorithms.

We proposed a two-stage process: (i) direction finding from \(K\) steered orientations, followed by (ii) distance recovery from a single beam-aligned power measurement. A full DEB and PEB analysis characterized identifiability (\(K\!\ge\!3\)) and showed the benefits of overdetermination (\(K\!\ge\!4\)). A GA-based orientation-set optimizer minimized the RMSE DEB across a 3D testbed and yielded practical steering patterns. Additionally, we introduce a ratio-based linearization of the Lambertian power model that yields closed-form GLS/WLS direction estimators for stage (i). We showed (Section~\ref{subsec:nr_independence}) that all three proposed direction estimators (GLS, WLS, NLS) are mathematically independent of the receiver orientation $\mathbf{n}_r$, which eliminates the need for receiver pose calibration during the direction-finding stage.

Monte Carlo simulations over $1{,}792$ testbed positions with $1{,}000$ trials confirmed the effectiveness of the proposed framework. For direction finding with $K=5$, all estimators achieved sub-degree mean angular error: GLS $0.54^\circ$, WLS $0.61^\circ$, NLS $0.52^\circ$, with the DEB at $0.46^\circ$. For full 3D positioning, the APE reached GLS $2.00$\,cm, WLS $2.25$\,cm, and NLS $1.90$\,cm at $K=5$, tracking the PEB ($1.43$\,cm) within a factor $\approx1.4\times$; increasing to $K=9$ improved GLS to $1.46$\,cm and NLS to $1.26$\,cm. Importantly, GLS/WLS reduce inference to a single $3\times3$ eigenproblem, yielding $\mu$s-level latency ($\approx0.025$\,ms), while NLS requires $\approx0.65$\,ms due to its iterative Levenberg--Marquardt optimization. Furthermore, under random receiver tilts (half-normal, $\sigma=5^\circ$, $\max=30^\circ$), all estimators degrade by less than $3\%$, confirming that receiver-orientation variations have negligible effect on direction-finding accuracy, as established analytically in Section~\ref{subsec:nr_independence}.

Beyond accuracy, direction finding constitutes the core enabling stage of the architecture. It requires no receiver rotation during the $K$ measurements, nor pose calibration, and the same periodic $K$-orientation scan can serve multiple receivers simultaneously. Once the direction $\hat{\mathbf{n}}_d$ is known, it enables cooperative beam alignment---steering the LED toward the PD and reorienting the PD toward the LED---which maximizes the link SNR for a subsequent distance-recovery measurement that completes the 3D position. Thus, beam-steering complexity is concentrated in the infrastructure (a single source). The full pipeline acquires $K+1$ sequential measurements ($K$ for direction finding and one for distance recovery). For $K=5$ with state-of-the-art steering mechanisms, the total acquisition window is on the order of milliseconds, resulting in sub-centimeter displacement at pedestrian speeds.

Future works include: (i)~experimental validation; (ii)~integration with beam-steered OWC links, where the direction-finding stage can be embedded into the beam-tracking loop and an intelligent refresh strategy can exploit the previous direction estimate to reduce the number of required orientations; and (iii)~extending the framework to optical integrated sensing and communication.

\appendices
\section{Proof of Equation \eqref{eq:beta_lin_gls}}
\label{app:beta_lin_gls}
For each orientation \(i\), the sample mean \(\hat\mu_i\triangleq\bar P_{r,i}\) satisfies
\(\hat\mu_i=\mu_i+n_i\) with \(n_i\sim\mathcal N(0,\sigma^2/N)\) mutually independent \cite{Kay1993}. 
For \(i\ge2\), define 
\(g(x_1,x_2)\!=\!(x_2/x_1)^{1/m}\) so that \(\hat\beta_i=g(\hat\mu_1,\hat\mu_i)\) and \(\beta_i=g(\mu_1,\mu_i)\).
A first–order multivariate Taylor expansion (delta method) of \(g\) at \((\mu_1,\mu_i)\) gives:
\begin{equation}
\hat{\beta}_i \;\approx\; \beta_i
+ \frac{\partial g}{\partial x_1}(\mu_1,\mu_i)\,n_1
+ \frac{\partial g}{\partial x_2}(\mu_1,\mu_i)\,n_i. 
\end{equation}
The partial derivatives of \(g\) are, for general \((x_1,x_2)\):
\begin{subequations}
\begin{align}
\frac{\partial g}{\partial x_1}(x_1,x_2) \;&=\; -\,\frac{1}{m}\,x_2^{\frac{1}{m}}\,x_1^{-\frac{1}{m}-1}
\;=\; -\,\frac{g(x_1,x_2)}{m\,x_1},\\[4pt]
\frac{\partial g}{\partial x_2}(x_1,x_2) \;&=\; \frac{1}{m}\,x_2^{\frac{1}{m}-1}\,x_1^{-\frac{1}{m}}
\;=\; \frac{g(x_1,x_2)}{m\,x_2}.
\end{align}
\end{subequations}
Evaluating at \((\mu_1,\mu_i)\) (so \(g(\mu_1,\mu_i)=\beta_i\)) gives:
\begin{subequations}
\begin{align}
\frac{\partial g}{\partial x_1}(\mu_1,\mu_i) \;&=\; -\,\frac{\beta_i}{m\,\mu_1},\\[2pt]
\frac{\partial g}{\partial x_2}(\mu_1,\mu_i) \;&=\; \frac{\beta_i}{m\,\mu_i}.
\end{align}
\end{subequations}
Hence:
\begin{equation}
\hat{\beta}_i \;\approx\; \beta_i
+ \frac{\beta_i}{m}\!\left(\frac{n_i}{\mu_i} - \frac{n_1}{\mu_1}\right),
\end{equation}
which is \eqref{eq:beta_lin_gls}.\hfill\(\square\)

\section{Proof of Equation \eqref{eq:cov_beta_gls}}
\label{app:cov_beta_gls}
From the first–order expansion (delta method):
\begin{equation}
\hat{\beta}_i \;\approx\; \beta_i
+ \frac{\beta_i}{m}\!\left(\frac{n_i}{\mu_i} - \frac{n_1}{\mu_1}\right),
\quad
n_\ell \sim \mathcal{N}\!\left(0,\tfrac{\sigma^2}{N}\right) \text{ i.i.d.},
\end{equation}
we can deduce that the ratio error is:
\begin{equation}
\tilde{n}_i \;\triangleq\; \hat{\beta}_i-\beta_i
\;\approx\; \frac{\beta_i}{m}\!\left(\frac{n_i}{\mu_i} - \frac{n_1}{\mu_1}\right).
\label{eq:tilden_i}
\end{equation}
Using $\operatorname{Var}(aX+bY)=a^2\operatorname{Var}(X)+b^2\operatorname{Var}(Y)$ for independent $X,Y$:
\begin{equation}
\operatorname{Var}[\tilde n_i]
=\Bigl(\tfrac{\beta_i}{m}\Bigr)^{\!2}\!
\Bigl(\tfrac{\sigma^2}{N\,\mu_i^{2}}+\tfrac{\sigma^2}{N\,\mu_1^{2}}\Bigr)
=\frac{\sigma^{2}}{N m^{2}}\,
\beta_i^{2}\!\left(\mu_i^{-2}+\mu_1^{-2}\right),
\label{eq:var_beta_gls_proof}
\end{equation}
which is \eqref{eq:var_beta_gls}. For $i\neq j$:
\begin{align}
\operatorname{Cov}[\tilde n_i,\tilde n_j]
&=\Bigl(\tfrac{\beta_i}{m}\Bigr)\!\Bigl(\tfrac{\beta_j}{m}\Bigr)
\operatorname{Cov}\!\left(\frac{n_i}{\mu_i}-\frac{n_1}{\mu_1},\,
                           \frac{n_j}{\mu_j}-\frac{n_1}{\mu_1}\right)\notag\\
&=\Bigl(\tfrac{\beta_i\beta_j}{m^2}\Bigr)\!
\left[0-0-0+\operatorname{Var}\!\Bigl(\tfrac{n_1}{\mu_1}\Bigr)\right]\notag\\
&=\frac{\sigma^{2}}{N m^{2}}\,
\beta_i\beta_j\,\mu_1^{-2},
\label{eq:cov_beta_gls_proof}
\end{align}
since $n_i,n_j,n_1$ are mutually independent and
$\operatorname{Var}(n_1/\mu_1)=\sigma^2/(N\mu_1^2)$.
Thus:
\begin{equation}
\operatorname{Cov}[\tilde{n}_i,\tilde{n}_j]
=
\frac{\sigma^{2}}{N m^{2}}\,
\beta_i\beta_j\,\mu_1^{-2}, \quad i\neq j,
\end{equation}
which is \eqref{eq:cov_beta_gls}.\hfill\(\square\)

\bibliographystyle{IEEEtran}
\bibliography{References}

\begin{IEEEbiographynophoto}{Kevin Acuna-Condori}
received the B.S. degree in electronical engineering from Universidad Nacional Mayor de San Marcos, Lima, Peru, in 2015, and the M.S. degree in mechatronics engineering from Pontifical Catholic University of Peru, Lima, Peru, in 2017. 
From 2018 to 2023, he led research activities at Cirsys, a company developing social robots and artificial intelligence for environmental applications.
Since 2024, he has been pursuing the Ph.D. degree at Université Paris-Saclay, Gif-sur-Yvette, France, with a focus on 3D visible light positioning.
His research interests include optical wireless positioning, machine learning, optical beam steering, and visible light communications.
\end{IEEEbiographynophoto}

\begin{IEEEbiographynophoto}{Bastien Béchadergue}
received the aeronautical engineering degree from ISAE-Supaero, Toulouse, France, the M.S. degree in communication and signal processing from Imperial College London, London, U.K., in 2014, and the Ph.D. degree in signal and image processing from UVSQ, Université Paris-Saclay, Gif-sur-Yvette, France, in 2017, for his work on visible light communication and sensing for automotive applications. From 2017 to 2020, he was in charge of the research activities at Oledcomm, one of the leading companies in the development of optical wireless communication products. Since 2020, he has been an Associate Professor with UVSQ, Université Paris-Saclay, where his research focuses on optical wireless communication and sensing.
\end{IEEEbiographynophoto}

\begin{IEEEbiographynophoto}{Hongyu Guan}
received the B.S. degree in electrical engineering from ENSEIRB, Talence, France, in 2007, and the Ph.D. degree in computer science from the University of Bordeaux I, Bordeaux, France, in 2012, for his work on embedded systems for home automation. He is currently a Research Associate and the Chief Project Engineer with the LISV laboratory, Versailles Saint-Quentin-en-Yvelines University, University of Paris-Saclay, Gif-sur-Yvette, France. His research interests include visible light communications, communication protocol, ubiquitous, data fusion, sensors, and nanometrology.
\end{IEEEbiographynophoto}

\begin{IEEEbiographynophoto}{Luc Chassagne}
received the B.S. degree in electrical engineering from Supelec, Gif-sur-Yvette, France, in 1994, and the Ph.D. degree in optoelectronics from the University of Paris XI, Orsay, France, in 2000, for his work in the field of atomic frequency standard metrology. He is currently a Professor with the LISV laboratory, University of Versailles, University Paris-Saclay, France. His research interests include nanometrology, sensors, and visible light communications.
\end{IEEEbiographynophoto}

\end{document}